\documentclass[12pt,reqno]{amsart}
\usepackage{graphicx,amsmath,amsfonts,amssymb,bbm,color}
\usepackage{psfrag}
\usepackage{epsf}
\usepackage{latexsym}
\usepackage{rotating}
\usepackage{soul}

\begin{document}
	
\title[Compactons in the sublinear KdV equation]{Stability and interaction of compactons \\in the sublinear KdV equation}
	
	\author[D.E. Pelinovsky]{Dmitry E. Pelinovsky}
\address[D.E. Pelinovsky]{Department of Mathematics, McMaster University, Hamilton, Ontario, Canada, L8S 4K1 and Institute of Applied Physics, Nizhny Novgorod, Russia} 
	
	\author[A.V. Slunyaev]{Alexey V. Slunyaev}
	\address[A.V. Slunyaev]{National Research University – Higher School of Economics, Nizhny Novgorod, Russia and Institute of Applied Physics, Nizhny Novgorod, Russia}
	
	\author[A.V. Kokorina]{Anna V. Kokorina}
\address[A.V. Kokorina]{Institute of Applied Physics, Nizhny Novgorod, Russia and Nizhny Novgorod State Technical University, Nizhny Novgorod, Russia}

		\author[E.N. Pelinovsky]{Efim N. Pelinovsky}
	\address[E.N. Pelinovsky]{National Research University – Higher School of Economics, Nizhny Novgorod, Russia;
		Institute of Applied Physics, Nizhny Novgorod, Russia and Nizhny Novgorod State Technical University, Nizhny Novgorod, Russia}

\begin{abstract}
Compactons are studied in the framework of the Korteweg--de Vries  (KdV) equation with the sublinear nonlinearity. Compactons represent localized bell-shaped waves of either polarity which propagate to the same direction as waves of the linear KdV equation. Their amplitude and width are inverse proportional to their speed. The energetic stability of compactons with respect to symmetric compact perturbations with the same support is proven analytically. Dynamics of compactons is studied numerically, including evolution of pulse-like disturbances and interactions of compactons of the same or opposite polarities. Compactons interact inelastically, though almost restore their shapes after collisions. Compactons play a two-fold role of the long-living soliton-like structures and of the small-scale waves which spread the wave energy.
\end{abstract}

\date{\today}
\maketitle

\section{Introduction}

{\em Compactons} are solitary waves of the finite length. They arise as the travelling waves in the degenerate Korteweg--de Vries (KdV) equation, 
\begin{equation}
\label{kdv}
\frac{\partial u}{\partial t} + \frac{\partial u^m}{\partial x} + \frac{\partial^3 u^n}{\partial x^3} = 0,
\end{equation}
where both the advection and dispersion terms are nonlinear with integer powers of $m \geq 2$ and $n \geq 2$. The degenerate KdV equation and the compactons were introduced and studied by Rosenau and Hyman \cite{RH93} (see also \cite{R94}). A variety of physical phenomena can be described by using the degenerate KdV equation (\ref{kdv}) and its modifications, e.g., the continuous limit of anharmonic oscillators in the elasticity theory \cite{RP05,RS05}, the magma dynamics \cite{SSW07}, the surface waves on vorticity discontinuities \cite{BH10}, sedimentation \cite{BBKT11}. A review of the degenerate KdV equation (\ref{kdv}) and its generalizations can be found in \cite{RZ18}.

Properties of compactons were studied extensively, such as duality between solitons and compactons \cite{OR96},  conservation laws \cite{V13}, Hamiltonian structure \cite{RZ17}, integrability \cite{HH20}, and stability \cite{Vlad}. Ill-posedness of the time evolution of the degenerate KdV equation due to loss of regularity and no continuous dependence on the initial data was discussed in \cite{WY13,ZR18}. Variational characterization of compactons and the proof of their stability were recently developed for a special version of the degenerate KdV equation in \cite{GHM20}; in this case, existence and uniqueness of solutions of time evolution (but not their continuous dependence on initial data) can be proven for positive initial data \cite{GHM19}.

Nonlinearity of the dispersion term ($n \geq 2$) is crucial for existence of the compactons in the degenerate KdV equation (\ref{kdv}). If $n = 1$, the travelling wave reduction of the generalized KdV equation can be integrated once and reduce to the second-order differential equation with a smooth vector field, for which existence and uniqueness of solitary waves  (homoclinic orbits) follows from the theory of differential equations. In this case, the second-order equation can further be integrated and solitary waves that decay to zero exponentially can be found in the analytic form \cite{Anco20}. It was claimed in \cite{W04a,W04b} (summarized in Section 13.12 of the book \cite{W09}) that compactons can be formally constructed for the modified KdV equation ($m = 3$, $n = 1$) and related equations by matching zero and nonzero solutions of the second-order differential equation; however, the claim is simply wrong because it contradicts the uniqueness theory for differential equation, e.g., the second-order derivative is not continuous across the breaking point. 

Besides the power nonlinearity in the advective term of the generalized KdV equation (with $n = 1$), the modular power nonlinearity of the form
\begin{equation}
\label{kdv-frac}
\frac{\partial u}{\partial t} + \alpha |u|^{\alpha - 1} \frac{\partial u}{\partial x} + \frac{\partial^3 u}{\partial x^3} = 0,
\end{equation}
has also been studied in literature, e.g., the Schamel equation in plasma physics with $\alpha = 1.5$ \cite{S73} or with $\alpha \in (1,2)$ \cite{MS06},  and the quadratic modular KdV equation with $\alpha = 2$ \cite{R13}. Another example is the modular KdV equation \cite{R16,NKR17} 
written in the form:
\begin{equation}
\label{kdv-modular}
\frac{\partial u}{\partial t} + \frac{\partial |u|}{\partial x} + \frac{\partial^3 u}{\partial x^3} = 0.
\end{equation}
The main model for our study is the sublinear KdV equation of the form
\begin{equation}
\label{kdv-sl}
\frac{\partial u}{\partial t} - \alpha |u|^{\alpha - 1} \frac{\partial u}{\partial x} + \frac{\partial^3 u}{\partial x^3} = 0,
\end{equation}
where $\alpha \in (0,1)$. This model was proposed 
by Rosenau in \cite{RosPRL}, where exact expressions for compacton solutions were obtained (see also \cite[Example 3]{RZ18}), as well as the conserved quantities. The first numerical simulations of compactons arising from a pulse-like initial condition were also reported in \cite{RosPRL}. 
Compactons in the sublinear Klein--Gordon and nonlinear Schr\"{o}dinger equations were further studied in \cite{Ros1,Ros2}.

The sublinear nonlinearity occurs in physical problems due to Bremsstrahlung radiation in plasma physics \cite{Rosenau1,Rosenau2} and in the modeling of 
the chemical reactions (autocatalysis) \cite{MN13,MN15}, for which the diffusion equation with the sublinear nonlinear terms is the main model. 
It was found long ago that such models lead to 
the finite-time extinction (vanishing of a solution after some finite time) 
\cite{Kalash,Martinson1,Kerner}. 

The sublinear KdV equation (\ref{kdv-sl}) with $\alpha \in (0,1)$ was derived 
in \cite{JP14,J17} in the theory of granular chains near the harmonic limit, for which the solitary waves has double-exponential decay to zero at infinity. 
Compactons in the sublinear KdV equation (\ref{kdv-sl}) represent 
the leading-order approximations of such fastly decaying solitary waves. 
The existence of compactons in the same model was recently reviewed 
in \cite{Salmoore}.

The previous preliminary results on the sublinear KdV equation (\ref{kdv-sl})  with $\alpha \in (0,1)$ call for the systematic study of the stability and robustness of compactons. {\em The main purpose of this work is to analyze stability of compactons and to explore propagation and interactions of compactons numerically.} The main results and organization of this article are explained as follows.

Existence of compactons expressed by the trigonometric functions is 
reviewed in Section~2. As is known from the previous work in \cite{RosPRL}, compactons are classical (three times continuously differentiable) solutions of the sublinear KdV equation if $\alpha \in \left(\frac{1}{3},1\right)$ 
and weaker (twice continuously differentiable) solutions 
for $\alpha \in (0,\frac{1}{3}]$.

Section~3 contains analytical results on the stability of compactons. 
We prove that compactons are local non-degenerate minimizers of energy 
under the fixed momentum in the space of symmetric compact functions 
with the same support for every $\alpha \in \left(\frac{1}{5},1\right)$. This result suggests energetic stability of compactons in the sublinear KdV equation. 
Moreover, we show under the same condition on $\alpha$ that the spectrum of linearized operator at the compacton solution is purely discrete.

In Section~4, we perform numerical simulations 
of single compactons, pulse-like initial disturbances (both wide and narrow), and interactions of two compactons of the same or opposite 
polarities in the particular case $\alpha = \frac{3}{4}$. Although interactions of compactons are similar to interactions of solitary waves, there are important differences related to generation of new compactons. These small-amplitude compactons replace the dispersive waves of the linear KdV equation, which do not propagate in the sublinear KdV equation. 

The summary of our results is given in the concluding Section~5.

\section{Compacton solutions} \label{Sec:CompactonSolutions}

Travelling waves are given of the form 
$u(x,t) = U(x+ct)$ with real-valued constants $c$. If $U(x) \to 0$ as $|x| \to \infty$ for solitary waves, integration of the third-order differential equation 
with zero integration constant yields the second-order equation 
\begin{equation}
\label{second-order}
U''(x) + c U(x) - |U(x)|^{\alpha - 1} U = 0.
\end{equation}
No solutions decaying to zero exist in (\ref{second-order}) for $c \leq 0$, 
hence the travelling solitary waves of the sublinear KdV equation (\ref{kdv-sl}) may only propagate to the left with $c > 0$.

Since $\alpha \in (0,1)$, the vector field in the system
$$
\frac{d}{dx} \left[ \begin{array}{c} U \\ V \end{array} \right] = 
\left[ \begin{array}{c} V \\ - c U + |U|^{\alpha -1} U \end{array} \right]
$$
is continuous at $U = 0$ but it is not Lipschitz continuous at $U = 0$. 
By the existence-uniqueness theory for differential equation, there exists 
a solution to the second-order equation (\ref{second-order}); however, the solution is not unique near $U = 0$. Therefore, it is allowed in some weak sense to concatenate nonzero solutions of the differential equaiton (\ref{second-order}) vanishing at some points 
with zero solutions, obtaining thus compactons, for which  functions 
$U$ are nonzero within an interval of a finite length denoted 
by ${\rm supp}(U) \subset \mathbb{R}$.

The sublinear KdV equation (\ref{kdv-sl}) is invariant under the scaling transformation
\begin{equation}
\label{scal-trans}
\tilde{u}(\tilde{x},\tilde{t}) = \gamma^{\frac{2}{1-\alpha}} u(x,t), \quad 
\tilde{x} = \gamma x, \quad \tilde{t} = \gamma^3 t,
\end{equation}
where both $u(x,t)$ and $\tilde{u}(\tilde{x},\tilde{t})$ satisfy the same equation and $\gamma > 0$ is arbitrary parameter. If ${\rm supp}(U)$ is a single interval, then its length can be normalized to $\pi$.
Since the sublinear KdV equation is also invariant with respect to translation in $x$, we can set 
$$
{\rm supp}(U) = [0,\pi].
$$

Integrating the second-order equation (\ref{second-order}) again  
with the zero integration constant yields the first-order quadrature 
in the form:
\begin{equation}
\label{first-order}
\frac{1}{2} (U')^2 + \frac{1}{2} c U^2 - \frac{1}{\alpha + 1} |U|^{\alpha + 1} = 0.
\end{equation}
The nonzero part of the compacton is found by integrating the first-order equation (\ref{first-order}). The exact solution on the normalized interval $[0,\pi]$ denoted by $U = U_1$ at $c = c_1$ is available in trigonometric functions \cite{RZ18,RosPRL}:
\begin{equation}
\label{compacton}
U_1(x) = \pm a_1 \sin^{\frac{2}{1-\alpha}}(x) \chi_{[0,\pi]}(x), \quad a_1 = \left[ \frac{(1 - \alpha)^2}{2 (1 + \alpha)} \right]^{\frac{1}{1-\alpha}}, \quad 
c_1 = \frac{4}{(1-\alpha)^2},
\end{equation}
where $\chi_{S}$ is the characteristic function on the interval $S \subset \mathbb{R}$ defined by $\chi_S(x) = 1$ if $x \in S$ and $\chi_S(x) = 0$ if $x \notin S$.

Using the scaling transformation (\ref{scal-trans}) with $\gamma = \pi/\lambda$, we find that the compacton $U(x)$ with amplitude $a$ and speed $c$ is located on the interval $[0,\lambda]$ with 
\begin{equation}
\label{scaling}
a = \left[ \frac{2}{(1+\alpha) c} \right]^{\frac{1}{1-\alpha}}, \quad \lambda = \frac{2\pi}{(1-\alpha) \sqrt{c}},
\end{equation}
where $c > 0$ is now arbitrary. It follows from (\ref{scaling}) that the amplitude $a$ and the length $\lambda$ decrease when the speed $c$ increases.

Closer to the left edge $x = 0$, the exact solution (\ref{compacton}) behaves like $U_1(x) \sim x^{\frac{2}{1-\alpha}}$. Since $\alpha \in (0,1)$, we have $U_1(0) = U_1'(0) = U_1''(0) = 0$, hence the compacton solution $U_1 \in C^2(\mathbb{R})$ is the classical solution 
of the second-order differential equation (\ref{second-order}) in spite of the breaking point at $x = 0$. Similar consideration holds near $x = \pi$ because $U_1(x)$ is symmetric about the point $x = \frac{\pi}{2}$. 

On the other hand, $U_1'''(x) \sim x^{\frac{3\alpha-1}{1-\alpha}}$, 
therefore, $U_1 \in C^3(\mathbb{R})$ is the classical solution 
of the sublinear KdV equation (\ref{kdv-sl}) if and only if 
\begin{equation}
\label{range-alpha}
\frac{1}{3} < \alpha < 1.
\end{equation} 

Due to the lack of uniqueness in the second-order equation (\ref{second-order}) at $U = 0$, one can combine compactons 
of the same and opposite polarities in order to construct bound states 
of compactons and anti-compactons propagating with the same speed, as well as the travelling  periodic waves with zero and nonzero mean values. 

If $\frac{2}{1-\alpha} = \ell$ is an integer (e.g. for $\alpha = \frac{1}{2}$ and $\ell = 4$), the periodic wave $U_1(x) = a_1 \sin^{\ell}(x)$ consists of finitely many Fourier harmonics. In general, the separation between nonzero parts of the compactons may be arbitrary and the Fourier series of the periodic waves may not be truncated at finitely many terms even if $\frac{2}{1-\alpha} = \ell$ is integer.

\section{Stability of compactons} \label{Sec:Stability}

Compactons can be obtained variationally, which is useful for understanding stability of compactons from the energetic point of view \cite{GHM20}. The sublinear KdV equation (\ref{kdv-sl}) formally admits conservation of three integrals, 
which correspond in classical mechanics to mass, momentum, and energy, respectively:
\begin{align}
\label{mass}
M(u) & = \int_{\mathbb{R}} u dx, \\
\label{momentum}
P(u) & = \frac{1}{2} \int_{\mathbb{R}} u^2 dx,
\end{align}
and
\begin{align}
\label{energy}
E(u) & = \frac{1}{2} \int_{\mathbb{R}} u_x^2 dx 
+ \frac{1}{1 + \alpha} \int_{\mathbb{R}} |u|^{1 + \alpha} dx.
\end{align}
The second-order equation (\ref{second-order}) is the Euler--Lagrange equation for the action functional
\begin{equation}
\label{action}
\Lambda_c(u) = E(u) - c P(u).
\end{equation}
so that $U$ is the critical point of $\Lambda_c(u)$. 
The energy $E(u)$ is not a $C^2$ functional at $u = 0$. 
Adding formally a perturbation $v$ to the compacton $U$ and 
expanding the action functional near $U$ give
\begin{equation}
\label{second-var}
\Lambda_c(U+v) = \Lambda_c(U) + \frac{1}{2} \langle L_{c} v, v \rangle + \cdots,
\end{equation}
where $\langle \cdot, \cdot \rangle$ is the standard inner product in $L^2(\mathbb{R})$ and $L_c$ is given by the differential expression
\begin{equation}
\label{Hessian}
L_c = -\partial_x^2 - c + \alpha |U(x)|^{\alpha - 1}.
\end{equation}

Without loss of generality, it is sufficient to consider the compacton $U = U_1$ given by (\ref{compacton}) for $c = c_1$, due to the scaling transformation (\ref{scal-trans}). In addition, we consider the class of compact 
perturbations $v$ with the support in ${\rm supp}(U_1)$ so that the operator $L_{c = c_1}$ is closed on the compact interval $[0,\pi]$ as the differential operator $L_1 : {\rm dom}(L_1) \subset L^2(0,\pi) \mapsto L^2(0,\pi)$ given by 
\begin{equation}
\label{self-adjoint}
L_1 := -\partial_x^2 - \frac{4}{(1-\alpha)^2} + \frac{2 \alpha (1+\alpha)}{(1-\alpha)^2 \sin^2(x)}, \quad x \in (0,\pi)
\end{equation}
and
\begin{equation}
\label{domain}
{\rm dom}(L_1) := \left\{ v \in L^2(0,\pi) : \quad L_1 v \in L^2(0,\pi) \right\}.
\end{equation}

We prove here that compactons $U$ are local non-degenerate minimizers of energy $E(u)$ subject to the fixed momentum $P(u)$ in the space of symmetric compact functions with the support in ${\rm supp}(U)$ for every 
\begin{equation}
\label{range-alpha-stab}
\frac{1}{5} < \alpha < 1.
\end{equation}  
Since both the energy and momentum are constants of motion, Lyapunov stability theory suggests that these compactons are energetically stable in the time evolution of the sublinear KdV equation (\ref{kdv-sl}).

In order to prove the statement, we first show that under the condition 
(\ref{range-alpha-stab}), the spectrum of the self-adjoint operator $L_1$ given by (\ref{self-adjoint}) and (\ref{domain}) is purely discrete and consists of real eigenvalues $\{ \mu_1, \mu_2, \mu_3, \dots \}$ such that 
\begin{equation}
\label{ordering}
	\mu_1 < \mu_2 = 0 < \mu_3 < \dots.
\end{equation}
Then, we show that the operator $L_1$ is strictly positive under the constraint of fixed momentum $P(u)$ and the spatial symmetry of perturbations $v$, so that 
$U_1$ is a local non-degenerate constrained minimizer of energy. 

In order to prove (\ref{ordering}), we consider the spectral problem $L_1 v = \mu v$ with the spectral parameter $\mu$. The spectral problem is defined by  the differential equation with regular singular points at $x = 0$ and $x = \pi$ (see \cite{Teschl}). For each regular singular point, indices 
of the Frobenius theory are given by roots of the indicial equation
\begin{equation}
\label{indicial-eq}
m (m-1) - \frac{2 \alpha (1+\alpha)}{(1 - \alpha)^2} = 0,
\end{equation}
from which we obtain 
\begin{equation}
\label{indices}
m_1 = \frac{1 + \alpha}{1 - \alpha}, \quad m_2 = -\frac{2\alpha}{1-\alpha}.
\end{equation}
Two fundamental solutions of the spectral problem $L_1 v = \mu v$ 
near $x = 0$ 
behave like $v_1(x) \sim x^{m_1}$ and $v_2(x) \sim x^{m_2}$ with $m_1 > 0$ and $m_2 < 0$. The second solution is not squared integrable near $x = 0$ if $m_2 < -\frac{1}{2}$ which yields the range (\ref{range-alpha-stab}). In this case, the end points $x = 0$ and $x = \pi$ are classified as the limit points by Weyl's theory and the spectrum of $L_1$ is defined by the admissible values of $\mu$ for which the two bounded solutions 
of the spectral problem $L_1 v = \mu v$ which behave like
$v(x) \sim x^{m_1}$ near $x = 0$ and $v(x) \sim (\pi - x)^{m_1}$ near $x = \pi$ intersect. This intersection is not generic and occurs at a countable number of isolated eigenvalues $\mu$; moreover,
each eigenvalue is obviously simple with only one linearly independent eigenfunction.

It remains to prove the ordering (\ref{ordering}) of eigenvalues. 
In order to compute the eigenvalues of $L_1$, we use the following substitution
\begin{equation}
\label{substitution}
v(x) = \sin^{m_1}(x) \phi(x),
\end{equation}
where $\phi$ solves the following second-order differential equation
\begin{equation}
\label{SL-equation}
- \phi''(x) - \frac{2 (1+\alpha)}{(1-\alpha)} \cot(x) \phi'(x) - \frac{(3+\alpha)}{(1-\alpha)} \phi(x) = \mu \phi(x).
\end{equation}
The first two eigenvalues and eigenfunctions are obtained from 
(\ref{SL-equation}) explicitly:
$$
\mu_1 = -\frac{3+\alpha}{1-\alpha}, \quad \phi_1(x) = 1
$$
and 
$$
\mu_2 = 0, \quad \phi_2(x) = \cos(x).
$$
By Sturm--Liouville theory, since $v_2(x) = \sin^{m_1}(x) \cos(x)$ has a single zero on $[0,\pi]$, $\mu_2$ is indeed the second eigenvalue of $L_1$
on $L^2(0,\pi)$, so that the ordering (\ref{ordering}) of simple eigenvalues of $L_1$ is proven.

Although the ordering (\ref{ordering}) was proven for $c = c_1$, 
the same ordering of eigenvalues holds for every $c > 0$ 
due to the scaling transformation (\ref{scal-trans}).
Thus, the compacton $U$ satisfying the second-order equation 
(\ref{second-order}) is a saddle point of $\Lambda_c$ with simple negative and zero eigenvalues of the quadratic form $\langle L_c v, v \rangle$ 
in the space of compact functions $v$ with ${\rm supp}(v) = {\rm supp}(U)$. 

Next, we show that the linear operator $L_c$ is strictly positive 
under the constraint of fixed momentum $P(u)$ and the spatial symmetry of perturbations.

The zero eigenvalue $\mu_2 = 0$ of the linear operator $L_c$ is related to the translational symmetry. If the perturbation $v$ is symmetric 
with respect to the middle point in ${\rm supp}(v) = {\rm supp}(U)$, then 
the zero eigenvalue is removed from the spectrum of $L_c$. 

Furthermore, the constraint of fixed momentum $P(u)$ imposes 
the constraint $\langle U, v \rangle = 0$ on the perturbation $v$. 
In order to show that the negative eigenvalue $\mu_1 < 0$ is removed 
under the constraint $\langle U, v \rangle = 0$, we need to show 
the validity of the Vakhitov--Kolokolov (slope) stability condition \cite{VK}:
\begin{equation}
\label{VK-criterion}
\langle L_c^{-1} U, U \rangle = \langle \partial_c U, U \rangle = \frac{d}{dc} P(U) < 0,
\end{equation}
where we have used that $L_c \partial_c U = U$ which follows from formal differentiation of the stationary equation (\ref{second-order}) in $c$.
It follows from (\ref{scal-trans}) and (\ref{scaling}) that 
$$
U(x) = \gamma^{-\frac{2}{1-\alpha}} U_1(\gamma x), \quad 
\gamma = \frac{1}{2} (1-\alpha) \sqrt{c}.
$$
Hence, we obtain 
$$
P(U) = \gamma^{-\frac{5-\alpha}{1-\alpha}} P(U_1),
$$
with $P(U_1) > 0$. Since $P(U) \sim \gamma^{-\frac{5-\alpha}{1-\alpha}}$ and $\gamma \sim c^{\frac{1}{2}}$, the slope condition (\ref{VK-criterion}) is satisfied for every $\alpha \in (0,1)$.

We summarize that the linear operator $L_c$ is strictly positive 
under the constraint of fixed momentum $P(u)$ and the spatial symmetry of perturbations $v$ so that the compacton $U$ is a local non-degenerate constrained minimizer of energy $E(u)$ in the space of symmetric compact 
perturtubations with the support in ${\rm supp}(U)$.

There are several loose ends in the rigorous stability theory 
for compactons in the sublinear KdV equation (\ref{kdv-sl}). 

First, well-posedness 
theory (e.g., continuous dependence on initial data) is out of reach 
in the time evolution of the sublinear KdV equation (see \cite{WY13} for analysis of well-posedness of another compacton equation). 

Second, existence of global minimizers of energy $E(u)$ under the fixed $L^2$ norm is difficult to prove because the subquadratic term of $E(u)$ is not controlled  
by Gagliardo--Nirenberg inequality (compared to the constraint on the $L^1$ norm considered in \cite{GHM20}). 

Third, since $E(u)$ is not $C^2$ at $u = 0$, the second variation 
in the expansion (\ref{second-var}) is related to the singular Sturm--Liouville operator $L_c$ with singularities at the end points of ${\rm supp}(U)$. 
This explains why the energetic stability has been obtained only in class of compact functions with the support in ${\rm supp}(U)$. Perturbations supported outside ${\rm supp}(U)$ do not belong to the positive subspace of energy because the linear operator $-\partial_x^2 - c$ is not positive for $c > 0$. These perturbations of the sign-indefinite energy may change the support of the compactons and may lead to  destabilization of the compacton $U$ in the time evolution. 

While the rigorous stability theory is opened for further studies, 
we give convincing numerical evidences of the 
stability and robustness of compactons in the time evolution 
of the sublinear KdV equation (\ref{kdv-sl}).

\section{Numerical simulation}

We have picked the power $\alpha = \frac{3}{4}$ which fits both the intervals (\ref{range-alpha}) and (\ref{range-alpha-stab}) for numerical study of the 
wave dynamics in the sublinear KdV equation (\ref{kdv-sl}). Simulations were run in a large spatial domain imposing periodic boundary conditions.

\subsection{Numerical method}

The split-step Fourier method (see, e.g., \cite{LoMei1985}) advances from one time level to the next one in two steps. The first step from $u(x,t)$ to $u_{NL}(x,t+\Delta t)$, is performed according to the nonlinear transport equation
\begin{align} \label{SSF_NonlinearStep}
	\frac{\partial u_{NL}}{\partial t} = \alpha |u_{NL}|^{\alpha-1}
 \frac{\partial u_{NL} }{\partial x}.
\end{align}
The transport equation (\ref{SSF_NonlinearStep}) 
is solved using the midpoint finite-difference approximation in two sub-steps,
\begin{align} \label{SSF_NonlinearSubSteps}
\left\{ \begin{array}{l} \displaystyle u_{NL}(x,t+\Delta t /2) = u(x,t) + \frac{\Delta t}{2} \frac{\partial}{\partial x} \left[ |u(x,t)|^{\alpha-1} {u(x,t)} \right], \\
\displaystyle
u_{NL}(x,t+\Delta t) = u(x,t) +  \Delta t \frac{\partial}{\partial x} \left[  |u_{NL}(x,t+\Delta t/2)|^{\alpha-1} u_{NL}(x,t+\Delta t/2)\right]. 
\end{array}\right.
\end{align}
Derivatives on the right-hand-side of (\ref{SSF_NonlinearSubSteps}) are calculated using the Fourier transform:
\begin{align}
	\frac{\partial f(x) }{\partial x} = 
	 \mathcal{F}^{-1} \left[ -i k \mathcal{F} \left(f(x) \right) \right],
\end{align}
where $\mathcal{F}(\cdot)$ denotes the discrete Fourier transform along the coordinate $x$, and $k$ is the corresponding array of wavenumbers.

The second step is performed in the Fourier domain using the analytic solution to the linear KdV equation:
\begin{align} \label{SSF_LinearStep}
	u(x,t+\Delta t)= \mathcal{F}^{-1} \left( e^{- i k^3 \Delta t} \mathcal{F} \left( u_{NL}(x,t+\Delta t) \right) \right) .
\end{align}

In our simulations, a better performance (with noise reduced) was demonstrated by a modification of the described above method, when the transport equation (\ref{SSF_NonlinearStep}) was solved using the Runge-Kutta method 
in the Fourier domain,
\begin{align} \label{SSF_NonlinearStepFur}
	\frac{\partial \hat{u}_{NL}}{\partial t} = 
	-ik \mathcal{F} \left(|u_{NL}|^{\alpha-1} u_{NL} \right), \qquad
	 \hat{u}_{NL} (k,t):= \mathcal{F} (u_{NL}(x,t)),
\end{align}
where the step from $\hat{u}_{NL}(x,t) = \hat{u}(x,t)$ to $\hat{u}_{NL}(x,t+\Delta t)$ is made using the standard fourth-order Runge-Kutta method. The subsequent solution (\ref{SSF_LinearStep}) is modified as follows,
\begin{align} \label{SSF_LinearStepMod}
	u(x,t+\Delta t)= \mathcal{F}^{-1} \left( e^{- i k^3 \Delta t} \hat{u}_{NL}(x,t+\Delta t) \right).
\end{align}

Sufficiently small time steps $\Delta t$  were chosen to achieve approximate conservation of the mass (\ref{mass}), momentum (\ref{momentum}), and energy (\ref{energy}) integrals. No spectral filters or wave smoothing which could damp the short-scale noise were applied.
In all simulations, fluctuations of the mass integrals did not exceed the level of $10^{-14}$, whereas fluctuations of the momentum and energy integrals were higher. Table~\ref{tab:Errors} presents the relative errors observed 
for the three conserved quantities in our numerical simulations.

\begin{table}[htp]
	\centering
\begin{tabular}{|c|c|c|c|}
	\hline
	& $|\Delta M/M|$  
	& $|\Delta P/P|$ 
	& $|\Delta E/E|$ \\
	\hline
	Fig.~\ref{fig:OneCompacton} 
	& $1.2 \cdot 10^{-15}$ 
	& $3.5 \cdot 10^{-10}$ 
	& $3.1 \cdot 10^{-10} $\\
	\hline
	Fig.~\ref{fig:PerturbedCompacton}
	& $2.3 \cdot 10^{-15}$ 
	& $6.0 \cdot 10^{-9}$
	& $3.2 \cdot 10^{-9}$ \\
	\hline
	Fig.~\ref{fig:BroadPulse}
	& $1.2 \cdot 10^{-15}$ 
	& $2.0 \cdot 10^{-11}$
	& $3.6 \cdot 10^{-8}$ \\
	\hline
	Fig.~\ref{fig:NarrowPulse}
	& $4.2 \cdot  10^{-15}$
	& $3.0 \cdot  10^{-7}$
	& $9.3 \cdot  10^{-5}$\\
	\hline
	Fig.~\ref{fig:SameSigns} 
	& $4.3 \cdot  10^{-15} $
	& $3.5 \cdot  10^{-7} $
	& $9.9 \cdot  10^{-8}$\\
	\hline
	Fig.~\ref{fig:CloseVelocities}
	& $3.2 \cdot 10^{-15} $
	& $2.2 \cdot 10^{-9} $
	& $1.4 \cdot  10^{-9}$\\
	\hline
	Fig.~\ref{fig:DifferentSigns}
	& $2.7  \cdot  10^{-15} $
	& $6.6  \cdot 10^{-7} $
	& $2.0  \cdot 10^{-7}$\\
	\hline
\end{tabular}
\caption{Relative errors in the conservation of the integrals (\ref{mass}), (\ref{momentum}), and (\ref{energy}) for the perfomed numerical simulations.}
\label{tab:Errors}
\end{table}

\subsection{Propagation of a single compacton}

The initial conditions for numerical simulations of compactons were based on the exact solution (\ref{compacton}) with (\ref{scaling}) and $\alpha = \frac{3}{4}$:
\begin{equation}
\label{CompactonIni}
u(x,0) = a \sin^8{\left(\frac{\pi x}{\lambda} \right)} \chi_{[0,\lambda]}(x), \quad
\lambda = \sqrt{56} \pi a^{1/8}, 
\end{equation}
where $a := \left(\frac{8}{7c}\right)^4$ is the amplitude parameter 
defined in terms of the speed $c > 0$. The edges of the compacton (\ref{CompactonIni}) are represented by relatively smooth functions which have the power-law dependence, $u(x,0) \sim x^8$ as $x \to 0$. Compactons of larger amplitude $a$ are wider in width $\lambda$ and move with smaller speed $c$.

\begin{figure}[htp]
	\centerline{\includegraphics[width=12cm]{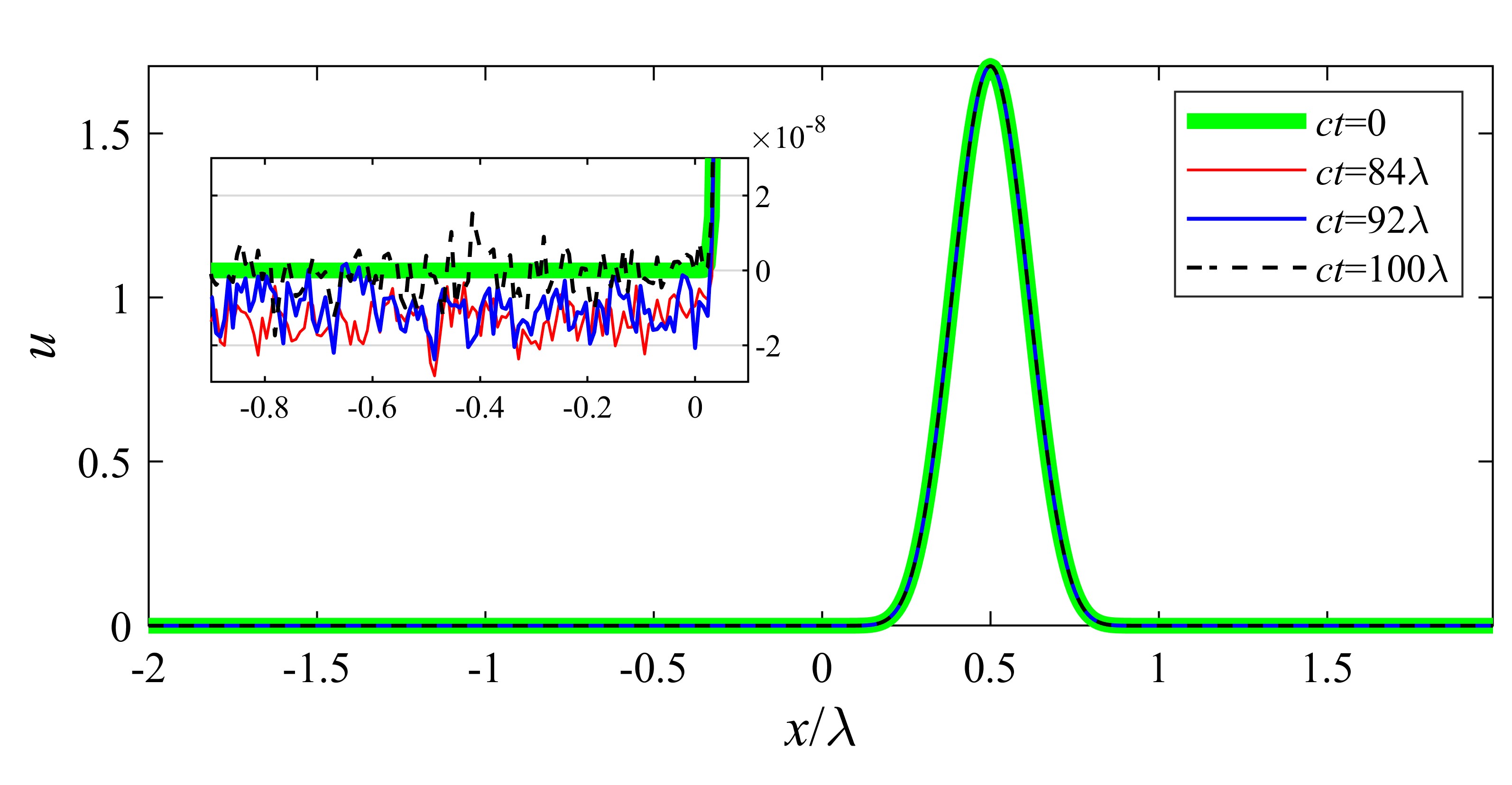}}
	\caption{Numerical simulation of a single compacton (\ref{CompactonIni}) with $c =1$ on the interval of the length $4\lambda$ subject to the periodic boundary conditions. The units of $x$ and $ct$ have been normalized by $\lambda$. }
	\label{fig:OneCompacton}
\end{figure}

Numerical simulation of a single compacton with the velocity $c = 1$ 
is shown on Fig.~\ref{fig:OneCompacton}. It follows from 
(\ref{CompactonIni})  that the compacton with $c = 1$ has 
$a \approx 1.7$ and $\lambda \approx 25.13$. The spatial domain has the total length $4\lambda$. It takes the time interval $T_{loop} = 4\lambda/c$
for the compacton to return to the initial position after moving along the periodic domain. 

Compacton (\ref{CompactonIni})  moves to the left during the time evolution.  Wave profiles after $21$, $23$, and $25$ cycles are superposed with the initial profile without any visible difference on Fig.~\ref{fig:OneCompacton}. The inset shows the small-amplitude radiation to the left of the compacton, from which it is clear that negligibly small short waves arise with time due to numerical errors. The short waves 
have a small negative mean value.

\subsection{Stability of a single compacton under small perturbations}

Here we study a respose of a single compacton to small perturbations of its width. The initial condition is taken in the form:
\begin{equation}
		\label{CompactonStrechted}
		u(x,0) =  a \sin^8{\left(\frac{\pi x}{\epsilon \lambda} \right)} \chi_{[0,\epsilon \lambda]}(x), \quad
		\lambda = \sqrt{56} \pi  a^{1/8}, 
\end{equation}
where $a > 0$ is the amplitude parameter and $\epsilon > 0$ is the deformation factor. The wave shape given by (\ref{CompactonStrechted}) coincides with the compacton (\ref{CompactonIni}) if $\epsilon=1$. If $\epsilon \neq 1$ but $|\epsilon - 1|$ is small, the support of the compacton is slightly perturbed.
	
\begin{figure}[htp]
	\centerline{\includegraphics[height=10cm]{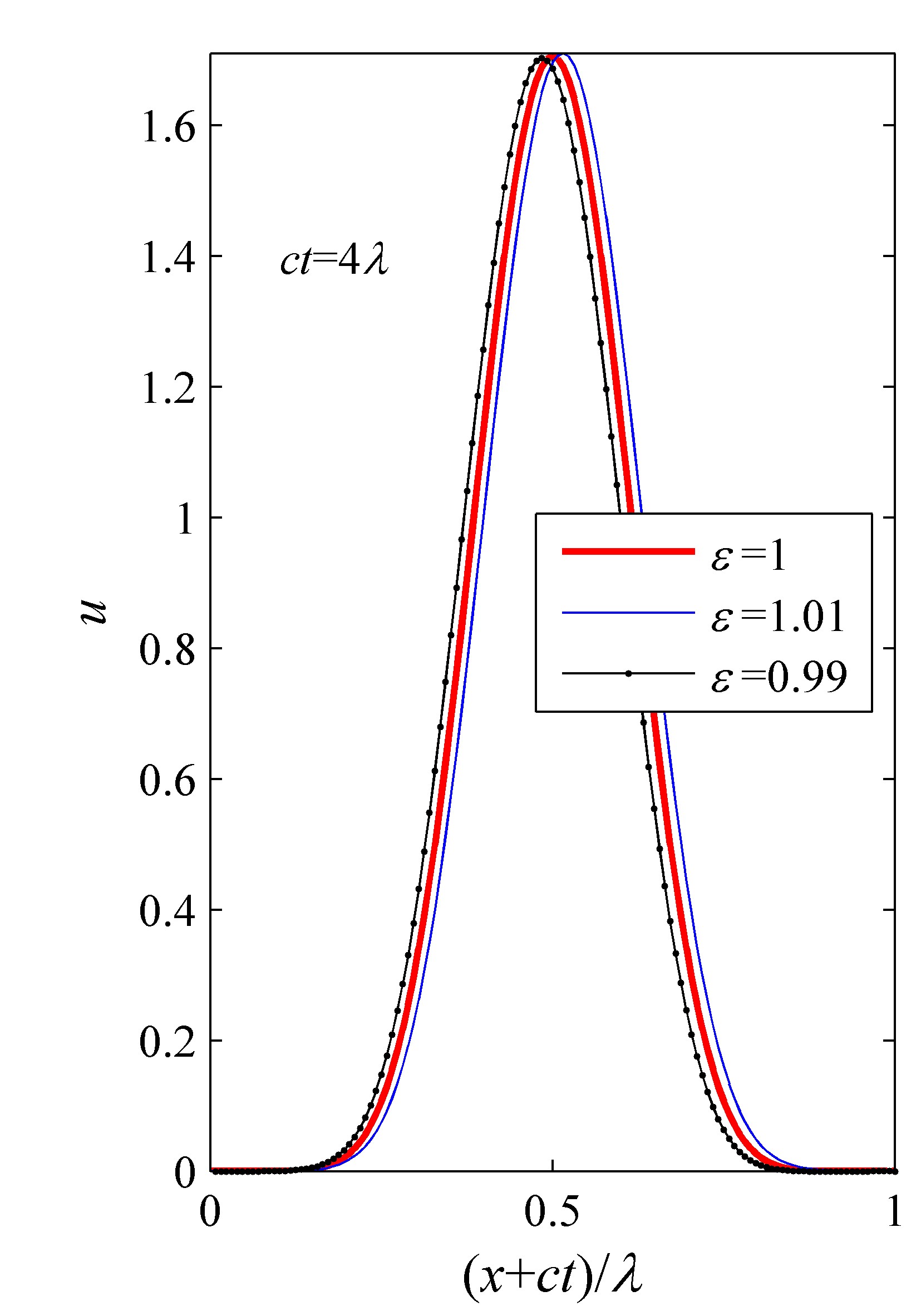}(a) 
	\includegraphics[height=10cm]{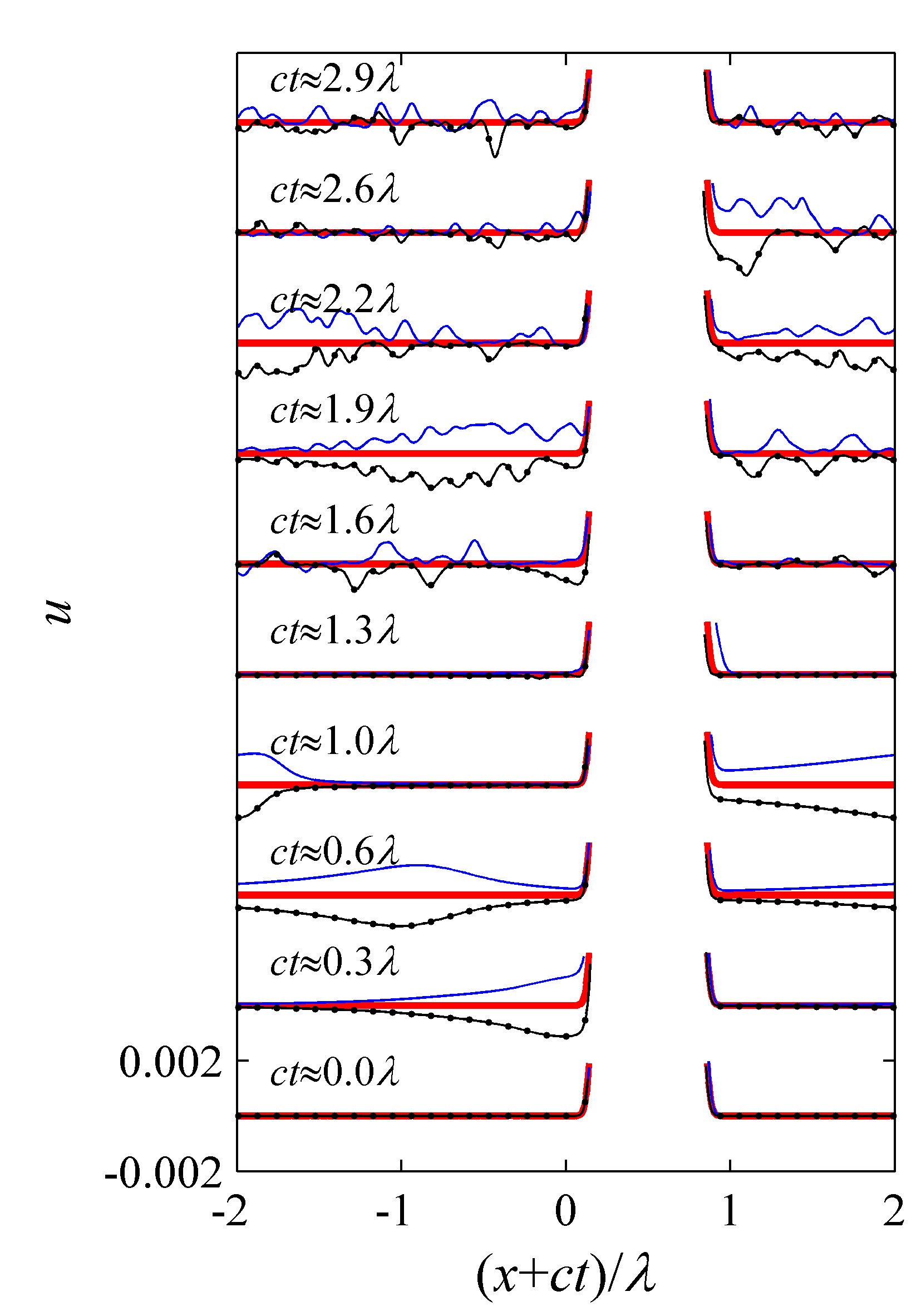}(b)}
	\caption{Evolution of compactons with perturbed supports in the periodic domain of the total length $4\lambda$: the wave shapes at the moment $ct=4\lambda$ (a) and the evolution of small-scale background waves (b).}
	\label{fig:PerturbedCompacton}
\end{figure}

Evolution of the initial conditions (\ref{CompactonStrechted}) with the same amplitudes $a=1.7$ (which corresponds to $c=1$) but with different widths, $\epsilon=0.99$ and $\epsilon=1.01$, is shown on Fig.~\ref{fig:PerturbedCompacton}. A slightly altered compacton is formed at a late stage, as shown on Fig.~\ref{fig:PerturbedCompacton}a. The longer support ($\epsilon = 1.01$) leads eventually to a slightly higher and slower compacton, while the shorter support ($\epsilon = 0.99$) causes the opposite effect. Evolution of the radiated background waves in the range of displacements $-0.002 \le u \le 0.002$ is shown in  Fig.~\ref{fig:PerturbedCompacton}b. Note that the solutions are plotted in the reference moving with the speed of the unperturbed compacton, $c$. 

At the first stage, the perturbed compactons emit long waves of different signs which propagate faster to the left. After some time, these waves re-appear from the right edge of the computational interval due to the periodic boundary condition. At the instant $ct\approx 1.3 \lambda$ the emitted waves enter the interval occupied by the compacton and seem to disappear. However, much less regular waves are radiated leftwards shortly after; they quickly acquire the chaotic character. Note that the amplitudes of the noisy background waves do not seem to grow in the course of the wave evolution.

Overall, the numerical simulations on Fig.~\ref{fig:PerturbedCompacton}, illustrate stability of a single compacton under a small perturbation of its support. Both computations for $\epsilon=0.99$ are $\epsilon=1.01$ are  qualitatively similar to each other. Though the energetic stability of compactons is proven for the symmetric perturbations with the same support, Fig.~\ref{fig:PerturbedCompacton} provides numerical evidences that compactons are stable even if the perturbations have different supports. 

\subsection{Transformation of pulse-like initial disturbances}
	
\begin{figure}[htp]
	\centerline{\includegraphics[width=12cm]{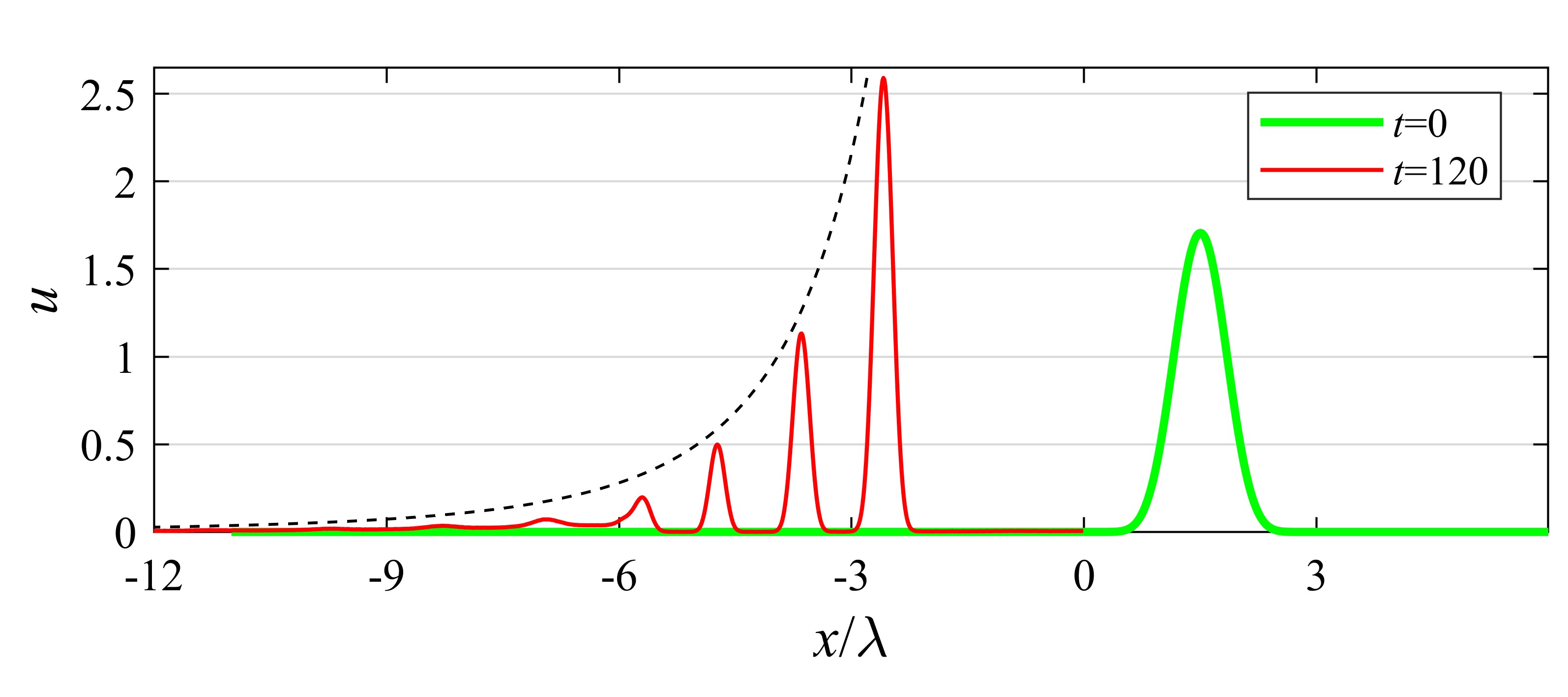}(a)}
	\centerline{\includegraphics[width=12cm]{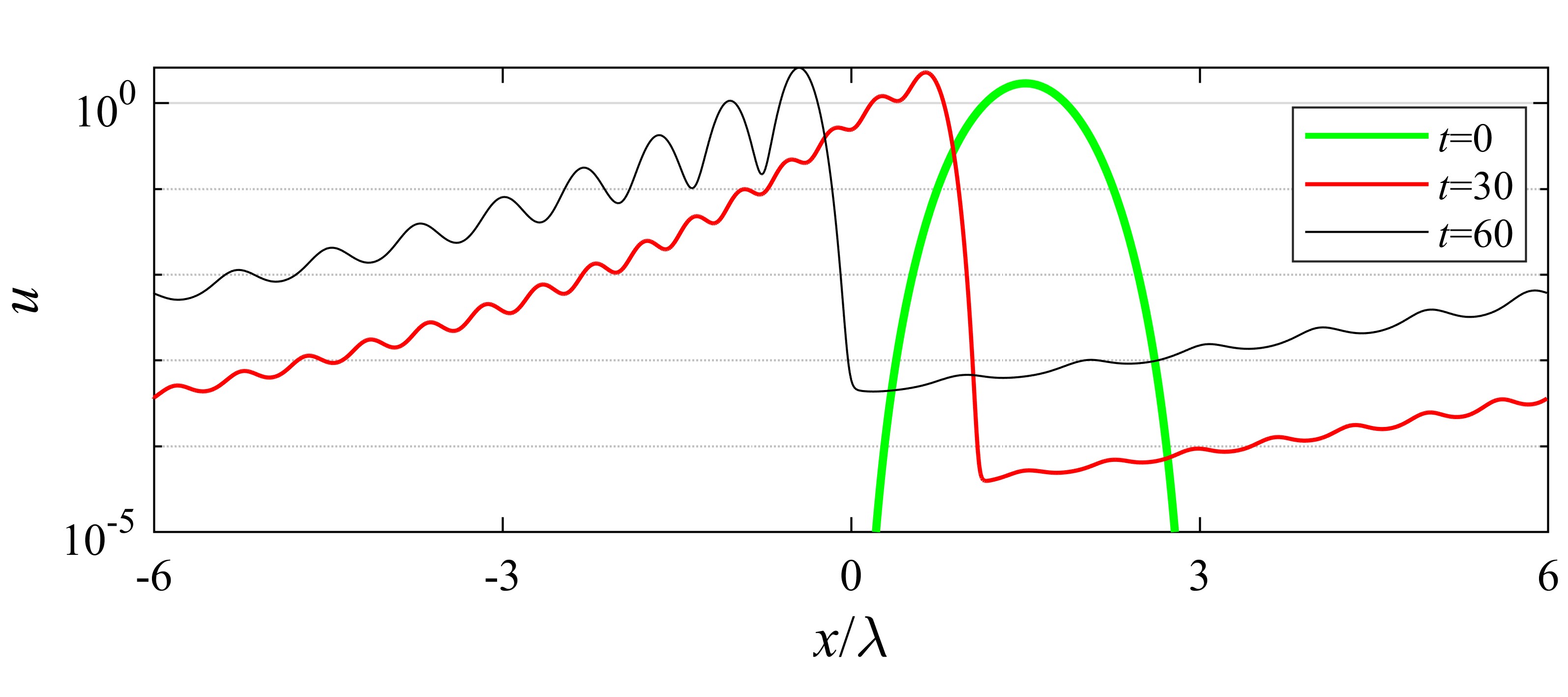}(b)}
	\caption{Evolution of a wide pulse-like initial perturbation: the long-time wave evolution (a) and the initial stage shown in semilogarithmic coordinates (b).}
	\label{fig:BroadPulse}
\end{figure}

\begin{figure}[htp]
	\centerline{\includegraphics[width=12cm]{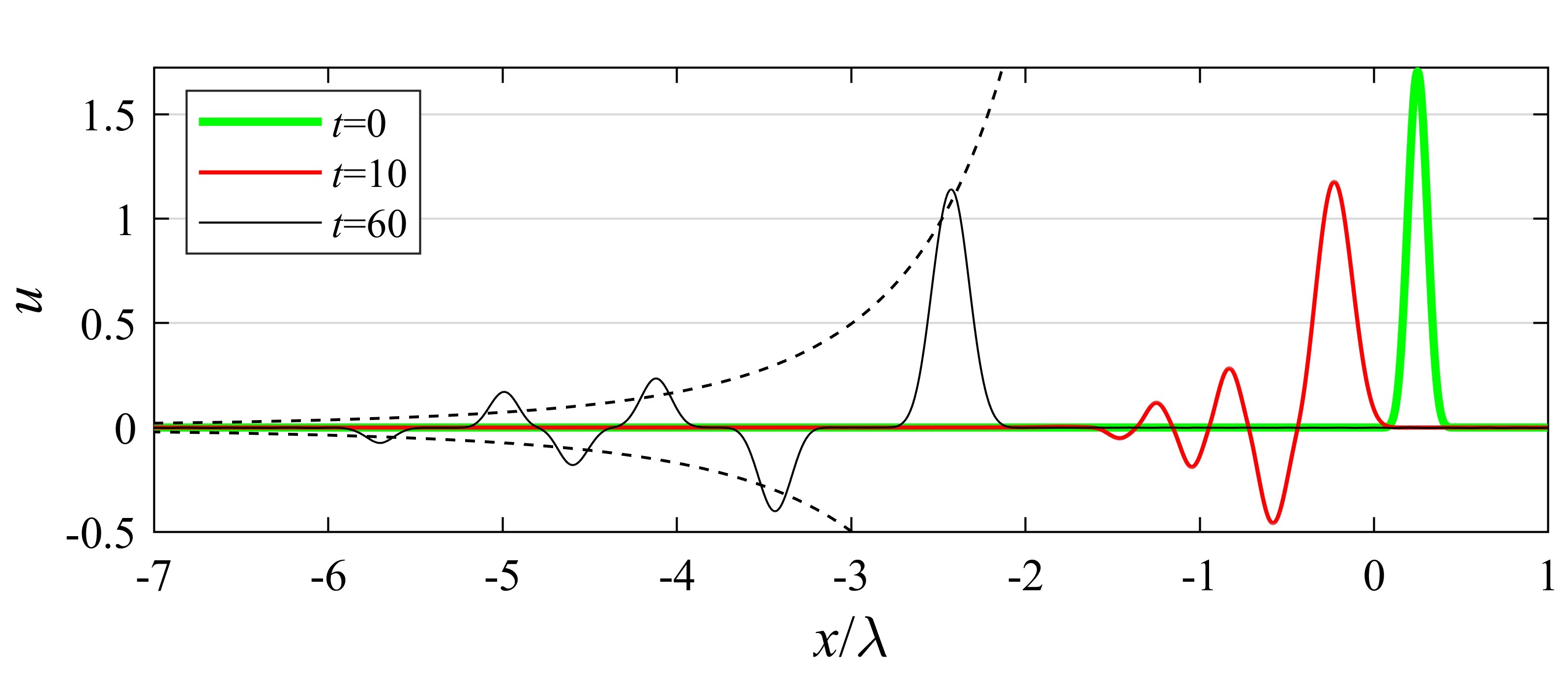}(a)}
	\centerline{\includegraphics[width=12cm]{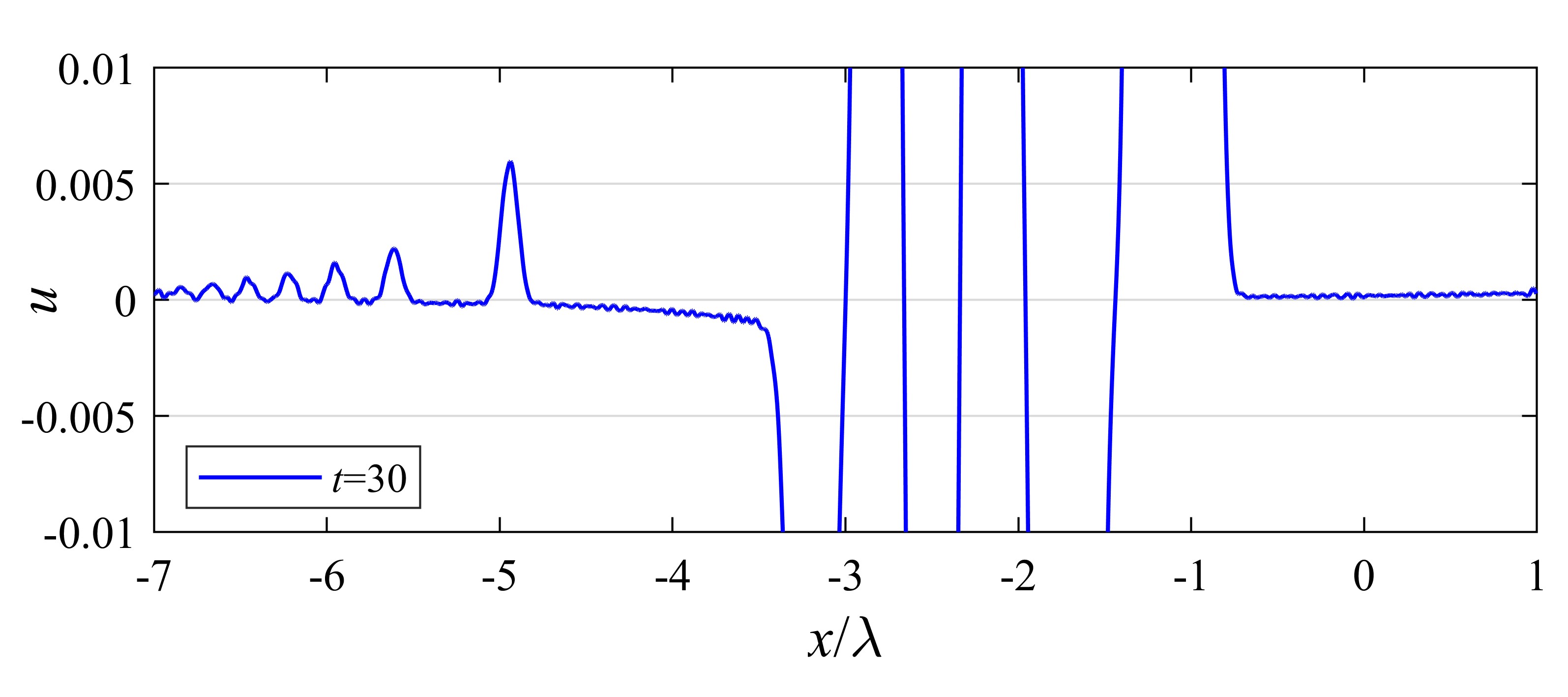}(b)}
	\caption{Evolution of a narrow pulse-like initial perturbation: the long-time wave  evolution (a) and the structure of the small-amplitude tail (b).}
	\label{fig:NarrowPulse}
\end{figure}

Here the initial conditions are taken in the form (\ref{CompactonStrechted}), but the parameter $\epsilon$ differs from $1$ significantly.
Figs.~\ref{fig:BroadPulse} and ~\ref{fig:NarrowPulse} 
show the outcomes of the time evolution for the stretched $(\epsilon = 3)$ 
and squeezed $(\epsilon = 0.5)$ initial condition (\ref{CompactonStrechted}) with $a = 1.7$, respectively.
The horizontal axes in the figures are scaled by the length $\lambda$ of this compacton.

\begin{figure}[htp]
	\centerline{\includegraphics[width=12cm]{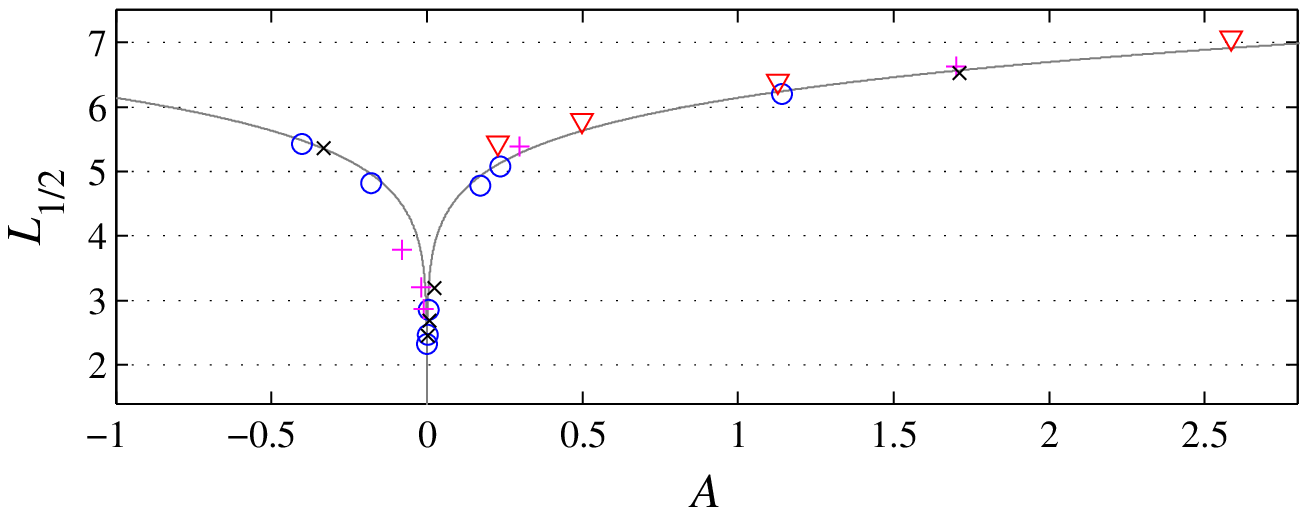}(a)}
	\centerline{\includegraphics[width=12cm]{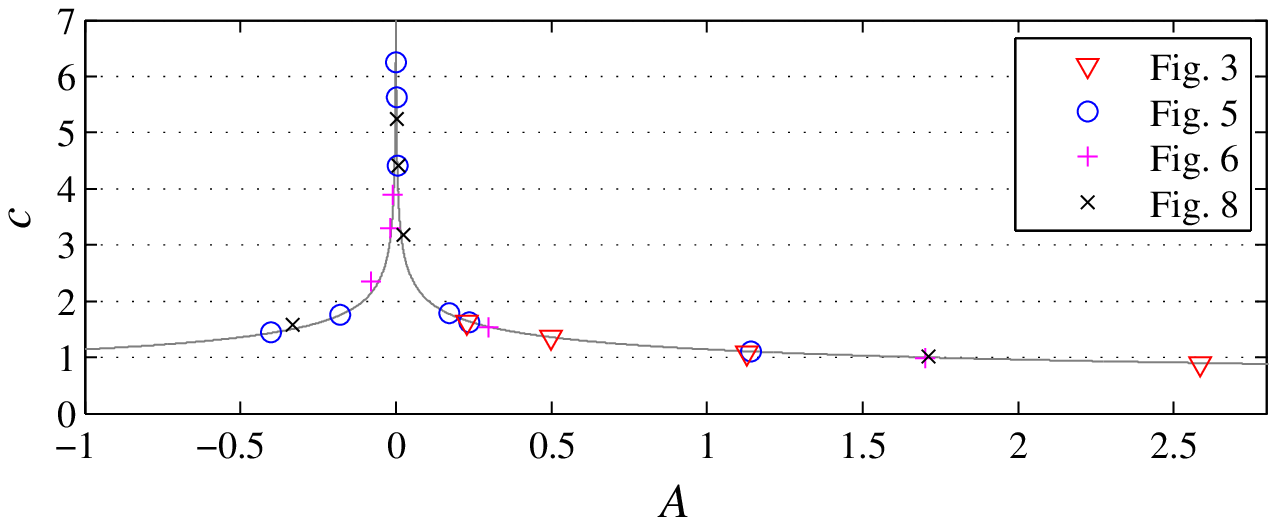}(b)}
	\caption{Parameters of the largest solitary waves descerned in the numerical simulations versus the relations which correspond to compactons.}
	\label{fig:SolitaryWaveParameters}
\end{figure}

For Fig.~\ref{fig:BroadPulse}, the initial condition (\ref{CompactonStrechted}) shown with the green thick line is thrice wider than a compacton of the same amplitude ($\epsilon = 3$). During the initial stage of the evolution the left slope of the hump moves leftwards and produces compactons as shown in Fig.~\ref{fig:BroadPulse}b in semilogarithmic axes. The distances between the emerging compactons increase with time. At time $t = 120$, 
it becomes apparent (Fig.~\ref{fig:BroadPulse}a) that the tail of the solution resolves into a train of compactons. 

The emergence of a train of compactons from the initial smooth data in the degenerate KdV equation (\ref{kdv}) was simulated previously, see e.g. the review \cite{RZ18}. The new compactons in our simulations are ordered according to the speeds, in the opposite order of their amplitudes. The amplitude of the highest compacton is significantly greater than the amplitude of the initial condition. The black dashed line in Fig.~\ref{fig:BroadPulse}a marks the locus of the compacton tops if they were propagating from the point of the initial condition with constant speeds $c$ defined by their final amplitudes $a$ by the formula 
\begin{equation}
c = \frac{8}{7 a^{1/4}}.
\label{speeds-amplitudes}
\end{equation}
According to Fig.\ref{fig:BroadPulse}a, compactons remain to the right of the dashed line, hence they experience deceleration when they are formed from the initial pulse-like perturbation.

We estimated the following parameters of the largest compactons which emerge in the evolution of a pulse-like initial disturbance: the amplitude $A$ at the local extremum (which may be either maximum of minimum), the width $L_{1/2}$ calculated at the levels $A/2$, and the velocity $c$. They are compared in Fig.~\ref{fig:SolitaryWaveParameters} with the theoretical dependences:
\begin{equation}
c = \frac{8}{7 |A|^{1/4}}, \quad 	L_{1/2} = 2 \sqrt{56} |A|^{1/8} \arccos{\left(2^{-\frac{1}{8}}\right)}.
	\label{width-amplitudes}
\end{equation}
Different symbols on Fig.~\ref{fig:SolitaryWaveParameters} correspond to different numerical simulations. The parameters of four largest solitary waves observed in the simulation shown in Fig.~\ref{fig:BroadPulse} are plotted in Fig.~\ref{fig:SolitaryWaveParameters}  with triangles. One may see that they approximately follow the theoretical curves. Some deviation from the theory may occur due to the discreteness of the data, but mainly due to the partial overlapping between the largest compactons and other background waves.

For  Fig.~\ref{fig:NarrowPulse}, the initial condition is twice narrower than a compacton of the same amplitude ($\epsilon = \frac{1}{2}$). During the early evolution, the transformed wave looks very much similar to the dispersive wave spreading described by the integrable KdV equation (see the curve for $t = 10$ in Fig.~\ref{fig:NarrowPulse}a). It becomes apparent during the later stage that new compactons with alternating signs arise during the time evolution. Similar to Fig.~\ref{fig:BroadPulse}, the dashed black curves indicate the expected locus of the compacton tops if they were propagating with constant speeds $c$ which depend on the amplitudes $a$ according to (\ref{speeds-amplitudes}). Compactons experience acceleration since the compactons are now to the left of the dashed curves. Parameters of the largest compactons are plotted in Fig.~\ref{fig:SolitaryWaveParameters} with circles.

In contrast to the simulation on Fig. \ref{fig:BroadPulse}, two sets of compactons emerge at the early stage of the evolution: a small number of compactons with relatively large amplitudes and alternating polarities and a sequence of much smaller positive compactons which are generated by the left slope of the evolving perturbation and quickly occupy the entire computational domain due to the high speed of small-amplitude compactons (see Fig.~\ref{fig:NarrowPulse}b). These small-amplitude compactons are not visible in Fig.~\ref{fig:NarrowPulse}a, however their amplitudes, widths and velocities approximately obey the relations for compactons (the parameters of three largest solitary waves are verified against the theory in Fig.~\ref{fig:SolitaryWaveParameters}).

Generation of compactons as a result of disintegration of pulse-like initial perturbations resembles much the scenarios of soliton generation in the classical KdV equation. However, compactons move to the left and their amplitudes are inverse proportional to their speeds according to (\ref{speeds-amplitudes}). The small compactons with decreasing amplitudes and increasing speeds play the role of dispersive waves in the linear KdV equation. 

\subsection{Interaction of compactons with same polarity}

Compactons propagate with different velocities if they have different amplitudes, and hence should eventually interact if a faster compacton is placed behind a slower one. In contrast to the case of solitary waves, the moment when two compactons start to interact, is defined exactly, when the boundaries of the compactons cross. 

\begin{figure}
	\centerline{\includegraphics[height=10cm]{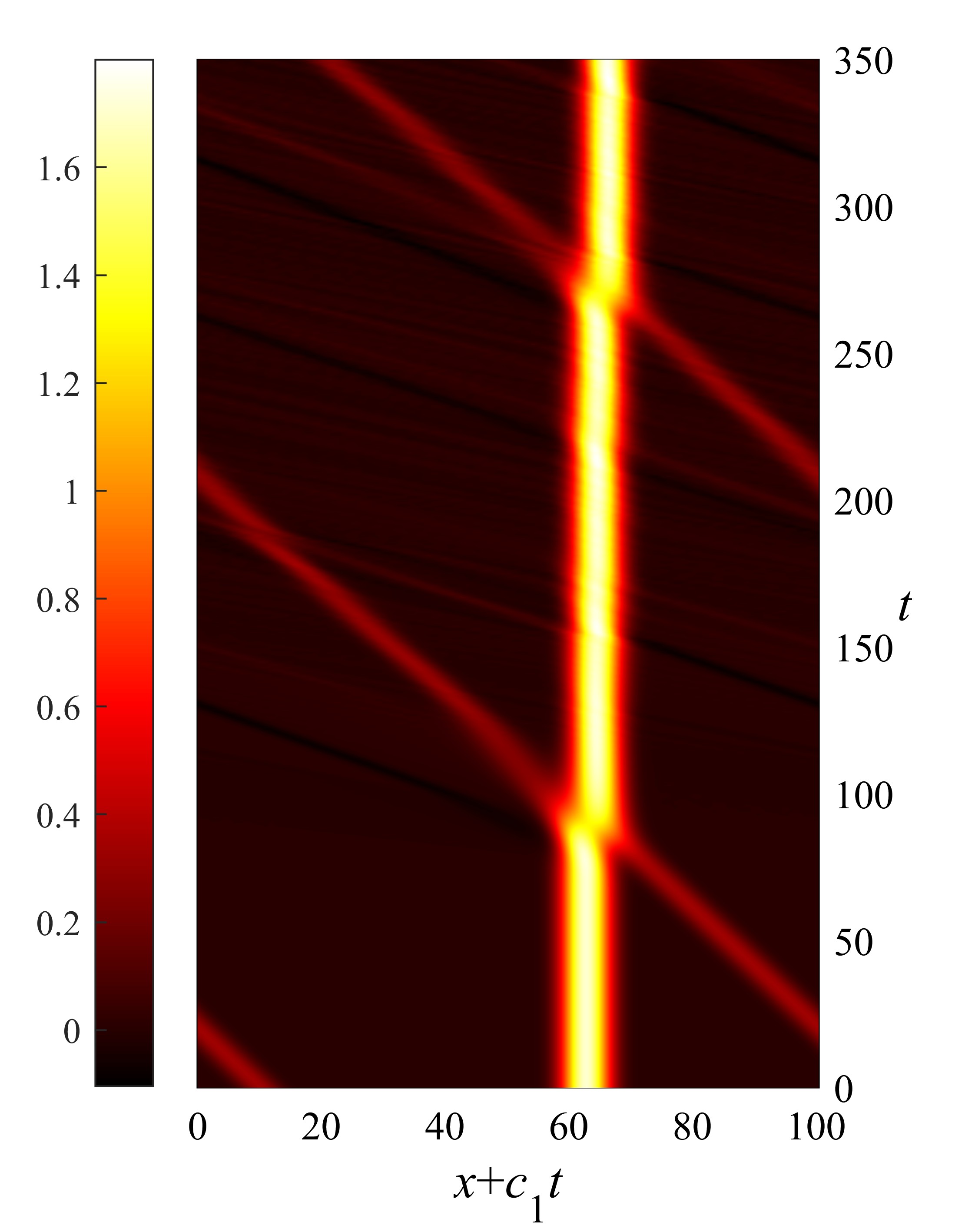}(a) 
	\includegraphics[height=10cm]{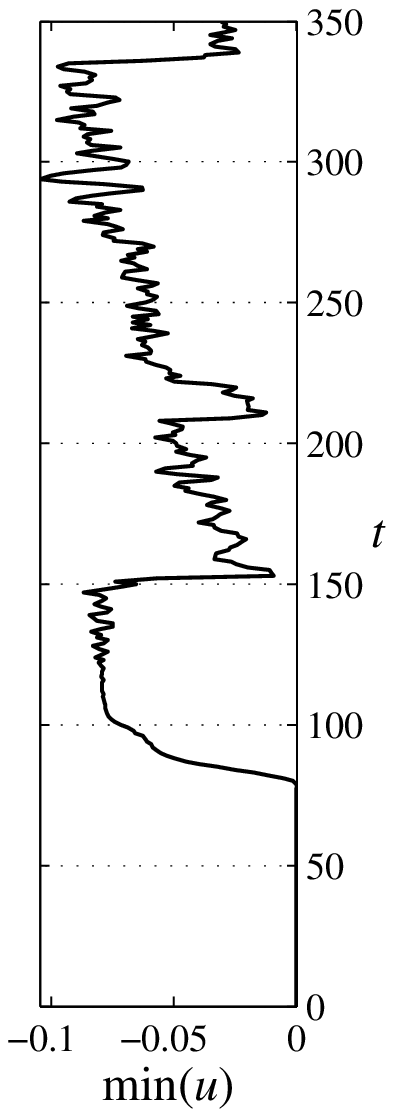}(b)
	\includegraphics[height=10cm]{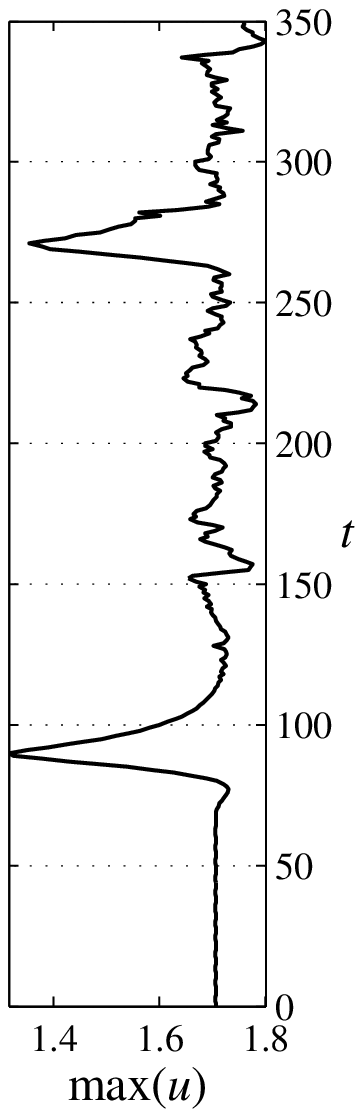}(c)}
	\centerline{\includegraphics[height=10cm]{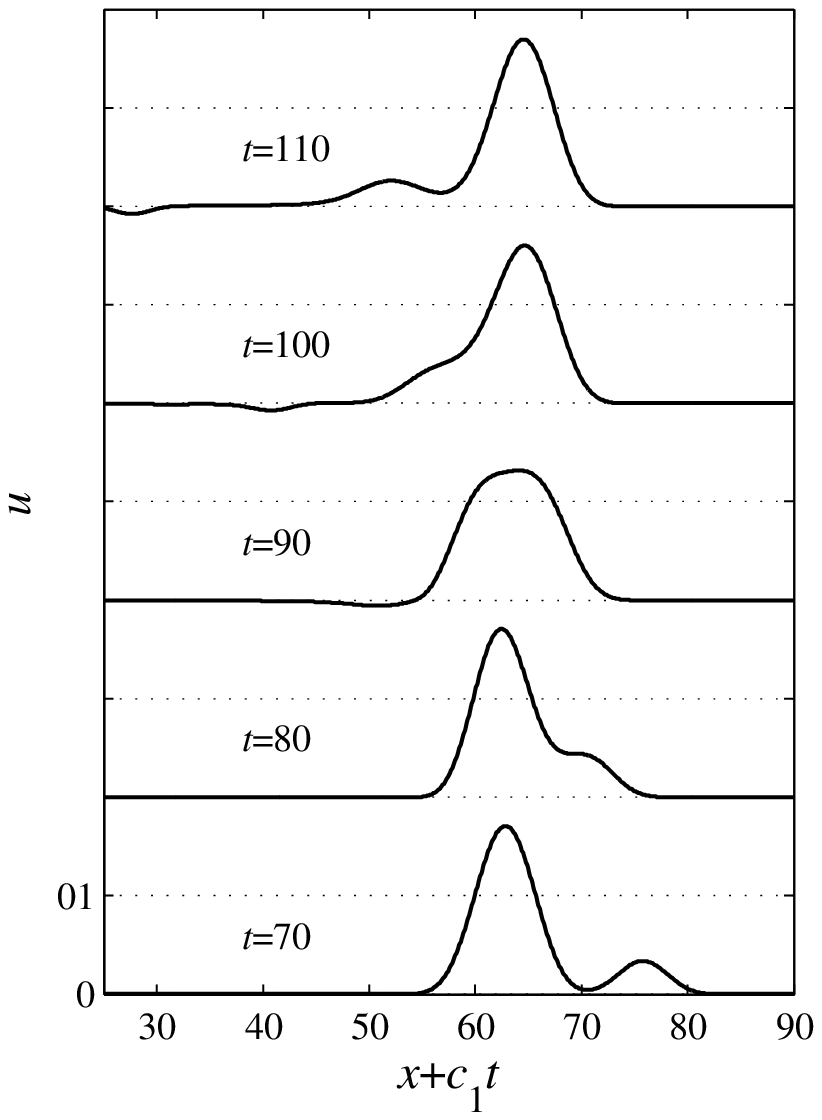}(d) 
		\includegraphics[height=10cm]{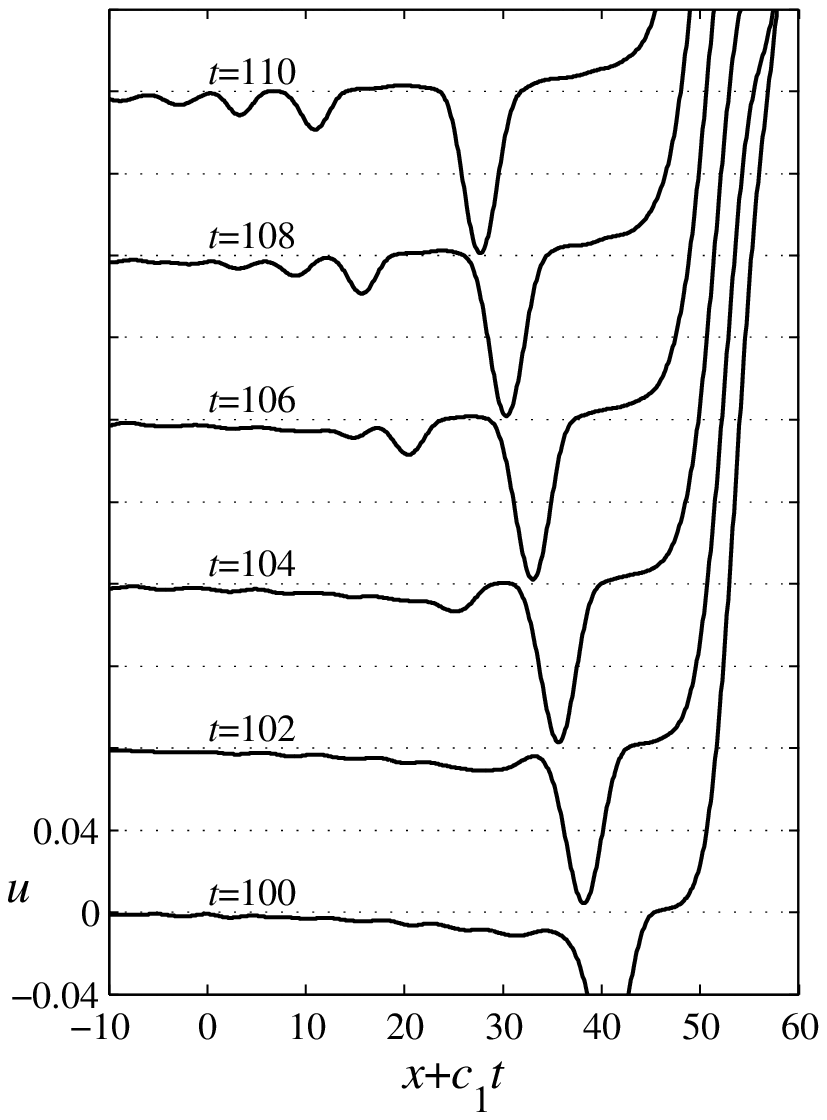}(e)}
	\caption{Numerical simulation of two compactons with the same polarity and different speeds $c_1 = 1$ and $c_2=1.5$: (a) the $(x,t)$ diagram, (b-c) evolution of $\min{u}$ and $\max{u}$ in time, and (d-e) snapshots during the first collision of the two compactons in different scales.}
	\label{fig:SameSigns}
\end{figure}

Fig.~\ref{fig:SameSigns} shows simulation of two compactons of the same polarity characterized by different speeds $c_1 = 1$ and $c_2 = 1.5$. At $t = 0$ the small compacton is located far to the left from the large one, though it propagates faster and catches up the other compacton thanks to the periodic boundary condition, see the $x$-$t$ diagram in Fig.~\ref{fig:SameSigns}a. The diagram is plotted in the reference moving with the speed $c_1$ of the slower (larger) compacton. The compactons do not exhibit any deformation until they encounter. The compactons collide at about $t \approx 90$ and seem to restore their shapes after the collision. Long after, they collide for the second time at $t \approx 270$ in a similar manner. The deceleration (shift to the right) of the slower (larger) compacton in the course of the collisions is obvious in Fig.~\ref{fig:SameSigns}a. A deeper investigation reveals that the faster (smaller) compacton experiences acceleration (shift to the left) during the collisions. 

A few snapshots of the solution during the phase of collision are shown in Fig.~\ref{fig:SameSigns}d. The wave is one-humped and is almost symmetric at $t = 90$. In general, the collision of compactons shown in Fig.~\ref{fig:SameSigns} looks very similar to the overtaking interaction of KdV solitons \cite{LeVegue,Kovalev}.

\begin{figure}
	\centerline{\includegraphics[height=10cm]{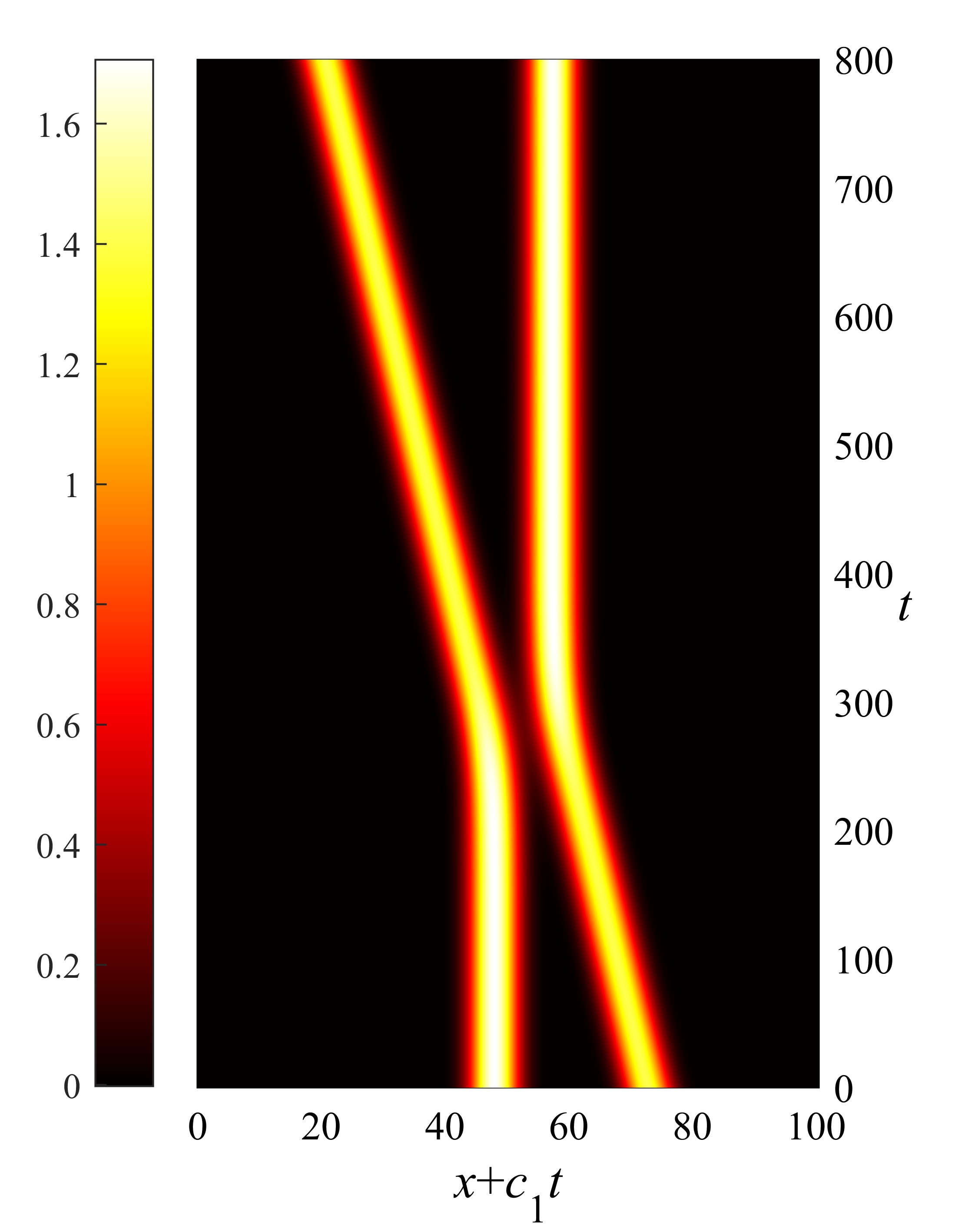}(a)
		\includegraphics[height=10cm]{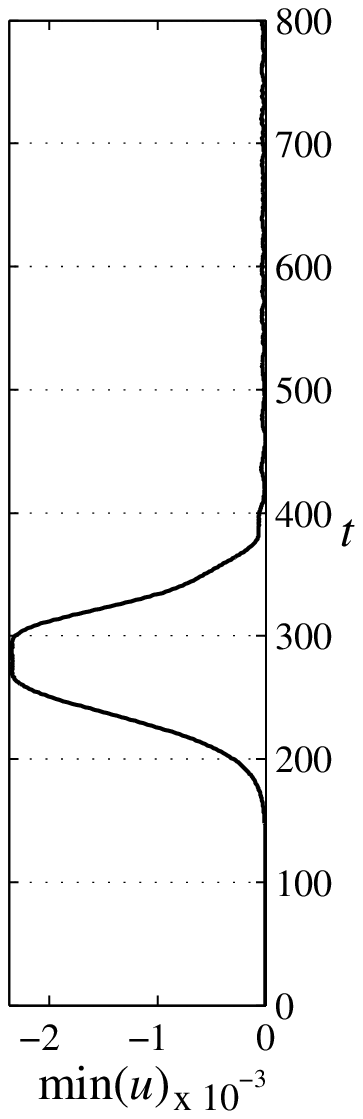}(b) \includegraphics[height=10cm]{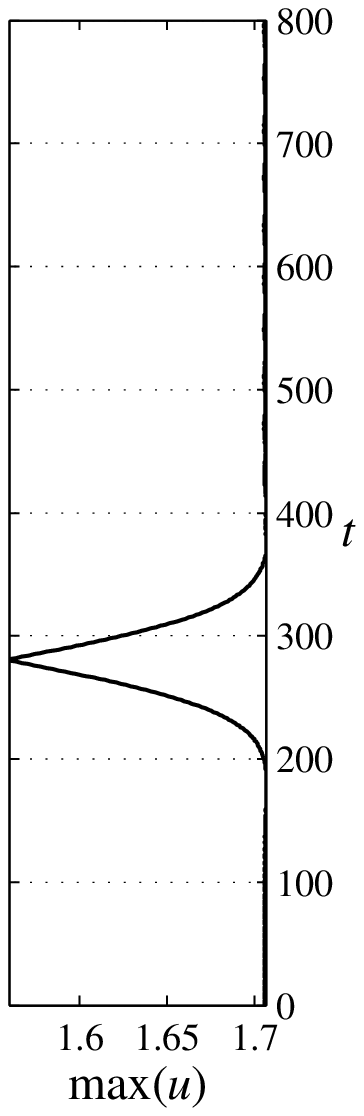}(c)}
	\centerline{\includegraphics[height=10cm]{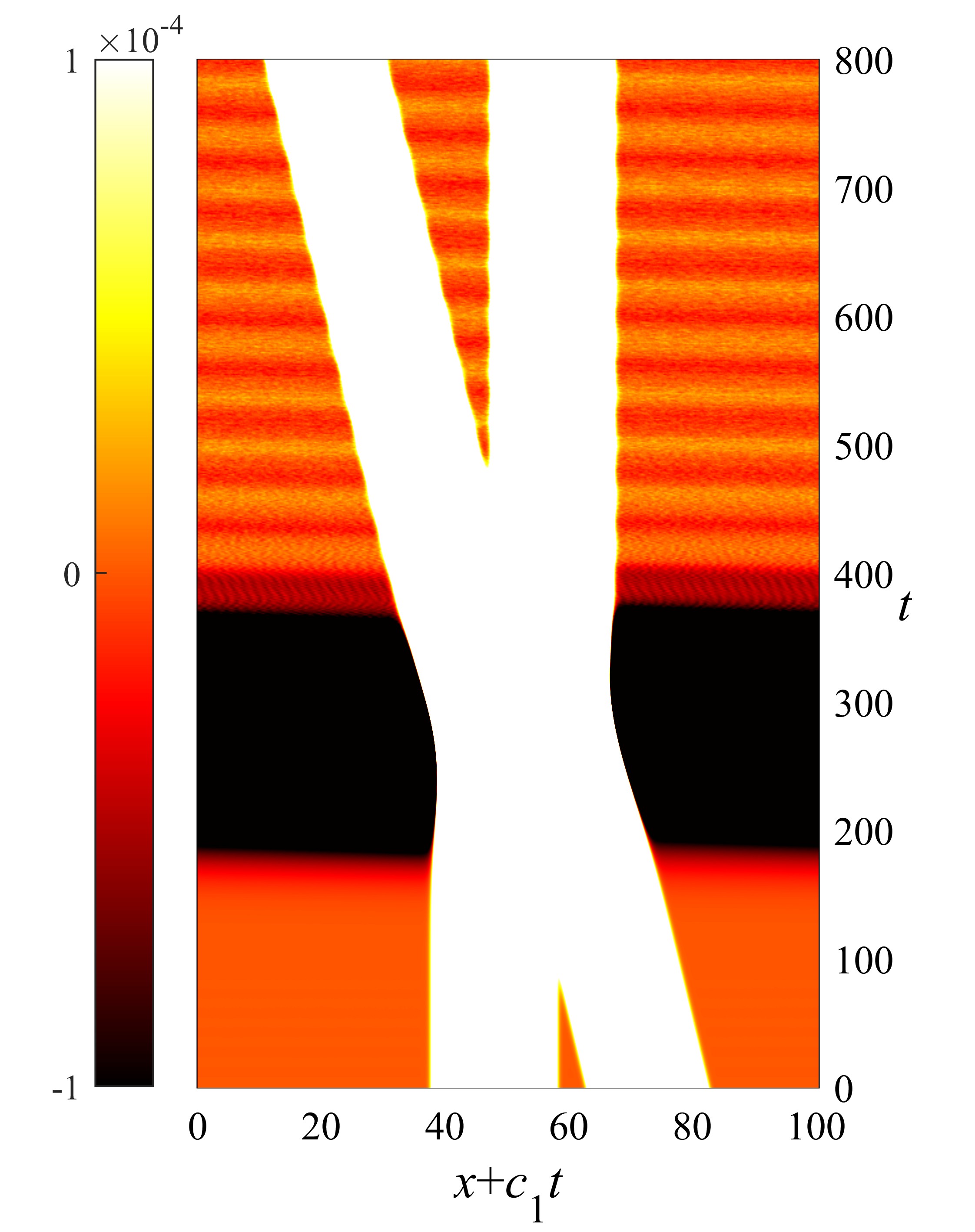}(d) 
		\includegraphics[height=10cm]{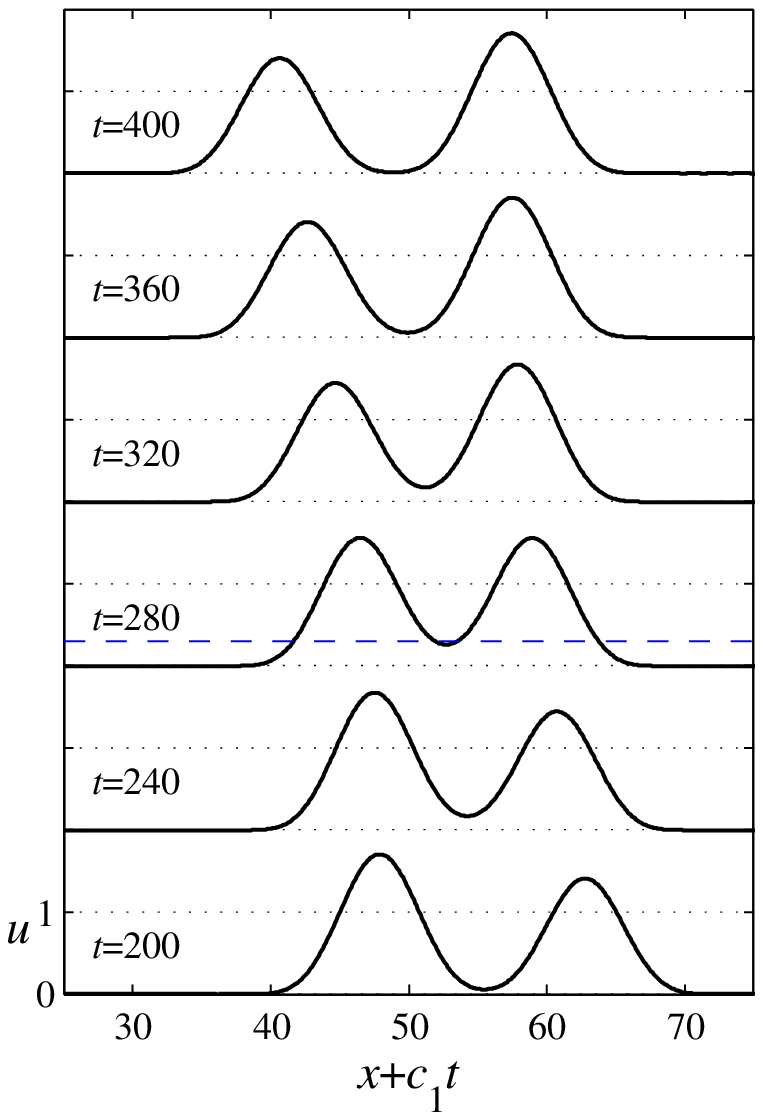}(e)}
	\caption{Numerical simulation of two interacting compactons with the same polarity and different speeds $c_1 = 1$ and $c_2 = 1.05$: (a) the $(x,t)$ diagram, (b-c) evolution of $\min{u}$ and $\max{u}$ in time, (d) the background waves and (e) the snapshots during the collision of two compactons.}
	\label{fig:CloseVelocities}
\end{figure}

The evolution of minimum and maximum of $u$ in the simulation domain is shown in Fig.~\ref{fig:SameSigns}b and Fig.~\ref{fig:SameSigns}c respectively. The constant value of $\max_x{u}$ before the collision is determined by the amplitude of the large compacton. When the compactons start to interact, the wave maximum shortly jumps up for a small amount, and then drops down more substantially. Later on, the wave maximum restores its original value, but only approximately. 

It was shown in \cite{SlyunyaevPelinovskii1999} and \cite{Slunyaev2001} for exact two-soliton solutions of the Gardner equation that when two interacting solitons form a symmetric waveshape, the field value in the middle of it is exactly the difference between the solitons' amplitudes, taking into account the amplitude signs. Solitons of the same polarity form smaller waves when collide, and solitons of the opposite polarities generate larger waves. This result was further generalized for an arbitrary number of interacting solitons within the modified KdV equation \cite{SlunyaevPelinovsky2016} and the Gardner equation \cite{Slunyaev2019}.
According to Fig.~\ref{fig:SameSigns}c, the minimum achieved during the amplitude drop is slightly smaller than the value $A_1 – A_2 \approx 1.4$.

The Gardner equation is a linear combination of the classic KdV and the modified KdV equations, and hence represents the general case of the integrable KdV equations in the form (\ref{kdv}) with $n=1$. The Gardner equation was also used in \cite{RosenauOron2014} as the reference example for the study of the wave dynamics within the degenerate KdV equation (\ref{kdv}) with $m=3$ and $n=3$.

Fig.~\ref{fig:SameSigns}e shows the compacton tails to the left of the interacting core (note different time instants in the plots in Fig.~\ref{fig:SameSigns}d and Fig.~\ref{fig:SameSigns}e). One may clearly see that the compactons interact inelastically. In the course of the collision they generate a series of waves of the opposite polarity, instead of dispersive waves of the linear KdV equation. The small-scale waves in Fig.~\ref{fig:SameSigns}e behave like compactons, the estimates of their amplitudes, widths and velocities approximately follow the curves for compactons in Fig.~\ref{fig:SolitaryWaveParameters}, see the magenta  pluses. The slowest negative compacton is large enough to be seen in Fig.~\ref{fig:SameSigns}d at $t = 100, 110$. The small negative compactons emerge from the depression wave and get ordered according to their speeds, as is seen on Fig.~\ref{fig:SameSigns}e at $t = 110$.

At longer time, more small-amplitude compactons of both signs are generated through the new collisions between compactons (see dark and bright traces in Fig.~\ref{fig:SameSigns}a), which result in irregular oscillations of the records of the wave minimum and maximum (Fig.~\ref{fig:SameSigns}b,c). Note, however, that these irregular oscillations do not show the tendency to grow in amount. The initially prescribed two compactons remain the most significant wave structures throughout the simulated time period.

Inelastic collisions of compactons, the emergence of new small-scale compactons (fission) including compactons of the opposite sign, and also the disintegration of one of interacting compactons were previously observed in the numerical simulations reported in \cite{RZ18,Vlad,RosenauOron2014} within other compacton-carying equations.

It is known from the classical KdV equation \cite{LeVegue,Kovalev} that solitons of the same polarity with close velocities repel, what leads to the second type of soliton interaction, of the exchange type. With the purpose to investigate this issue, the simulation of two positive compactons with close velocities is performed, and the exchange interaction of compactons is found for $c_1 = 1$ and $c_2 = 1.05$, see Fig.~\ref{fig:CloseVelocities}. The graphs in Fig.~\ref{fig:CloseVelocities}a show how the two compactons exchange their amplitudes and undergo coordinate shifts similar to the case displayed in Fig.~\ref{fig:SameSigns}. A few shapes of the interacting compactons are plotted in Fig.~\ref{fig:CloseVelocities}e which resemble the exchange interaction of KdV solitons \cite{LeVegue}; the waves do not form a single-humped wave. A symmetric wave appears in the course of the interaction at $t \approx 280$, which corresponds to the local minimum of the function $\max_x{u}$. The deepest point of the hollow between the humps approximately corresponds to the value $A_1 – A_2$ (shown with the horizontal dashed line in Fig.~\ref{fig:CloseVelocities}e for the solution at $t =280$). 

A detailed investigation of the picture of interaction of compactons with close velocities reveals new effects shown on Fig. \ref{fig:CloseVelocities}d. The small relative velocity of the compactons yields the long period of nonlinear interaction, thus the compactons might change significantly. However, no generation of small-scale waves with amplitudes appreciably greater than the level of the numerical noise is observed. Instead, a wave set-down in the entire simulation domain is produced by the colliding compactons during the period $200 \lesssim t \lesssim 400$, which may be seen in Fig.~\ref{fig:CloseVelocities}d. This figure replicates the surface displayed in Fig.~\ref{fig:CloseVelocities}a, but the displacements are shown with the color in a much smaller range of magnitudes, $1\cdot10^{–4}$  (see the colorbar). The white traces in the figure correspond to the interacting compactons, while the black area –- to the set-down. The evolution in time of the set-down amplitude may be seen in the record of $\min_x{u}$ in Fig.~\ref{fig:CloseVelocities}b. It is small in magnitude, but is well above the level of noise of the code. 

The constant set-down vanishes when the compactons detach, and the wave maximum nearly returns to the value of the initial condition. However, at this stage the background starts to oscillate with a small amplitude. The oscillations become apparent through the almost horizontal color stripes in Fig.~\ref{fig:CloseVelocities}d. Besides, tiny oscillations of the functions $\min_x{u}$ and $\max_x{u}$ in Figs.~\ref{fig:CloseVelocities}b,c for $t > 400$ reveal a new wave component which emerges in the course of the collision. The strips in the $x$-$t$ diagram in Fig.~\ref{fig:CloseVelocities}d possess a small inclination which corresponds to perturbations propagating to the left. However, the amplitude of the oscillations is not much larger than the accuracy of the simulation, thus the investigation of this dynamics requires more efforts and lies beyond the scopes of the present work. 

\begin{figure}
	\centerline{\includegraphics[height=10cm]{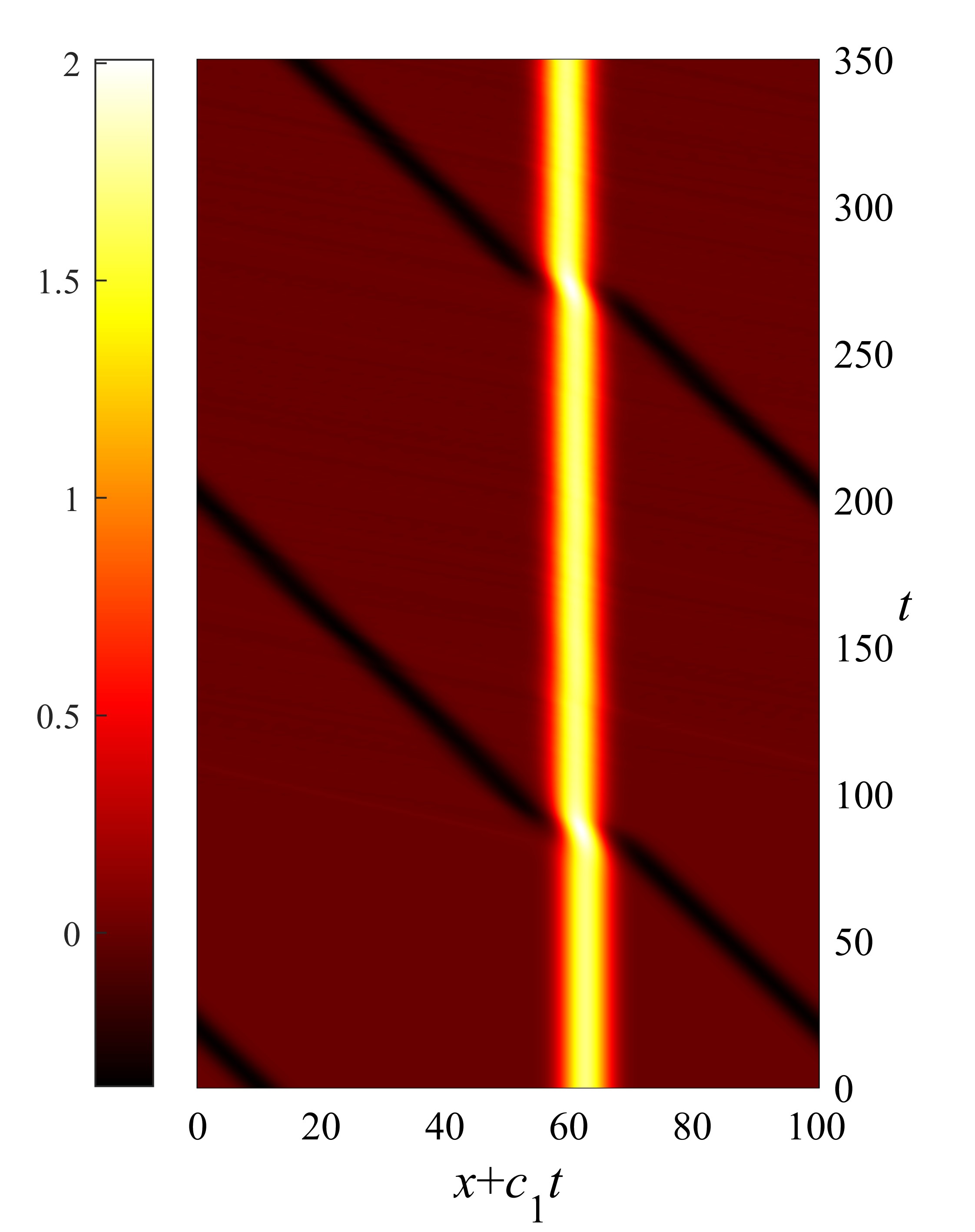}(a)
		\includegraphics[height=10cm]{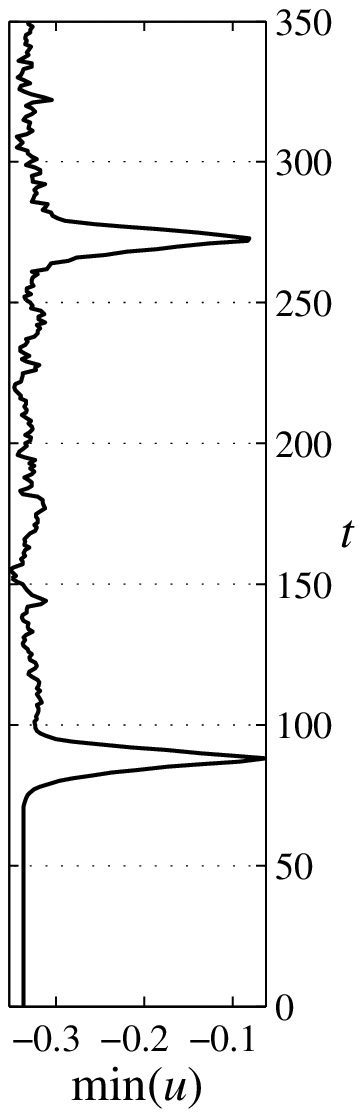}(b) \includegraphics[height=10cm]{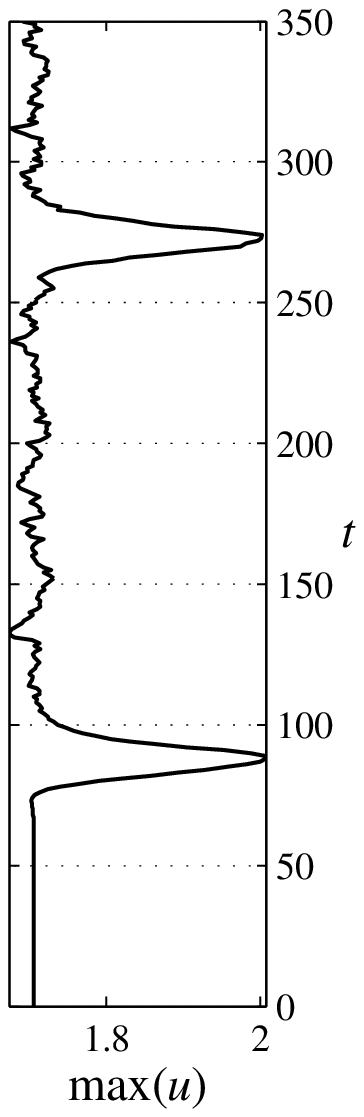}(c)}
	\centerline{\includegraphics[height=10cm]{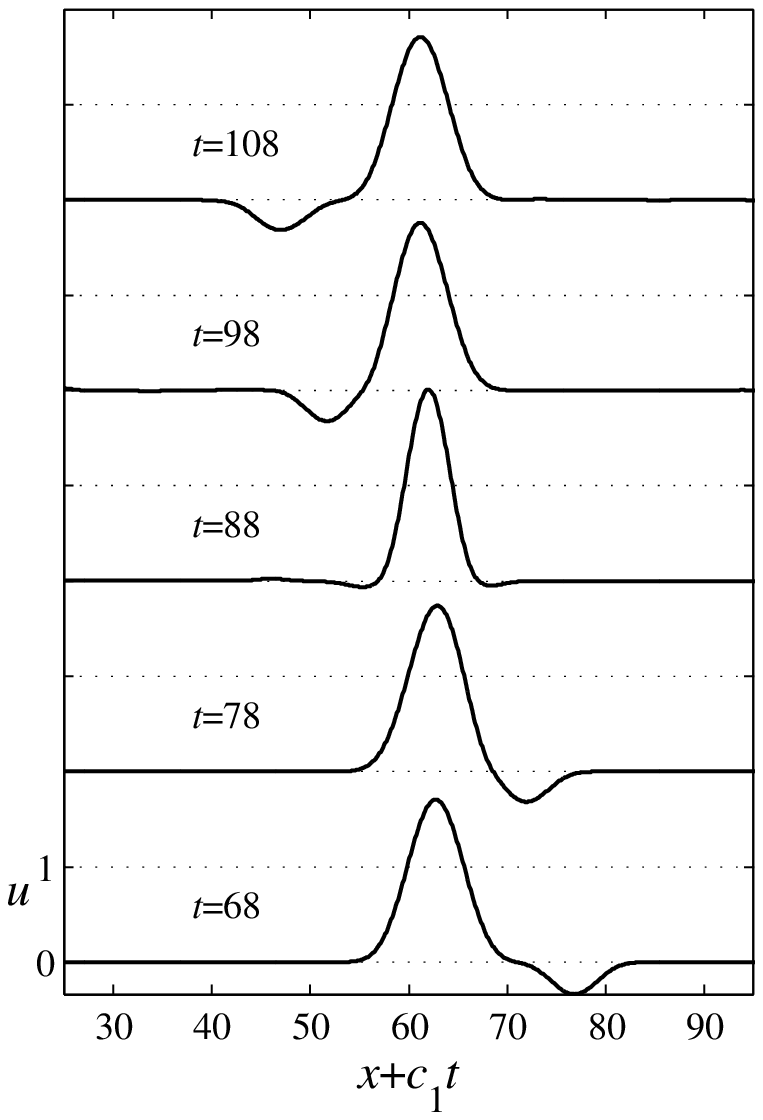}(d) 
		\includegraphics[height=10cm]{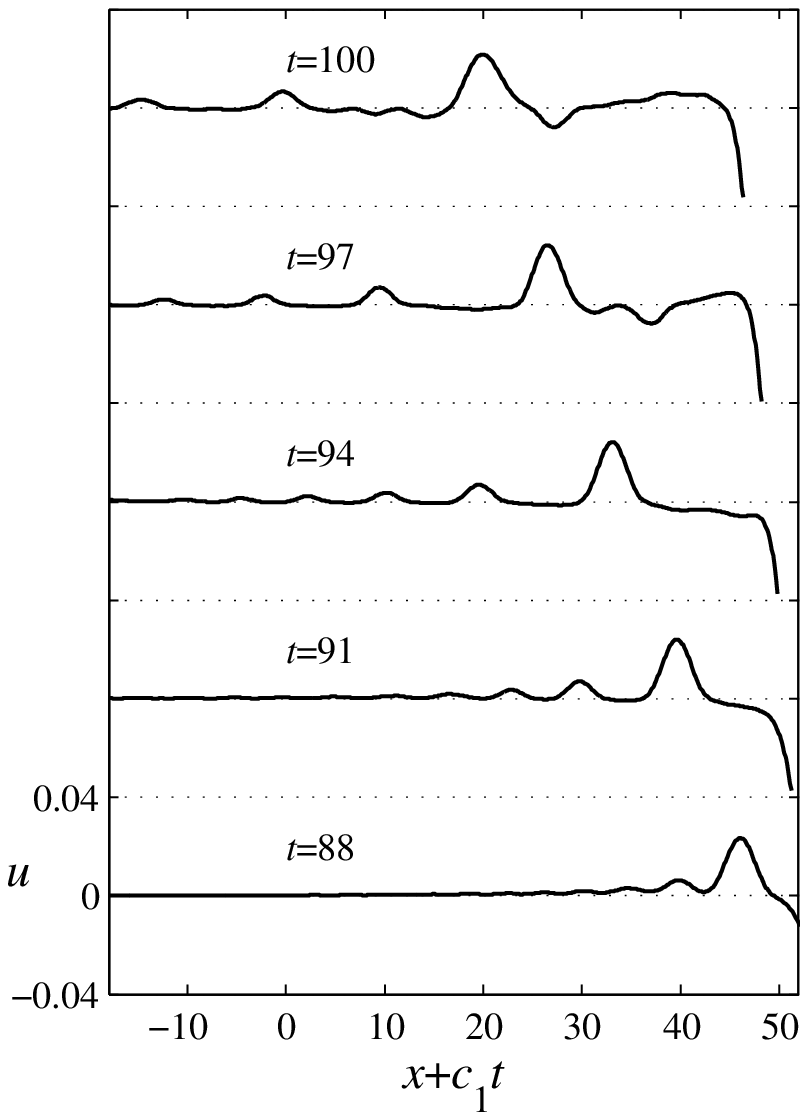}(e)}
	\caption{The same as Fig.~\ref{fig:SameSigns} but for two interacting compactons with opposite polarities and different speeds $c_1 = 1$ and $c_2 = 1.5$.}
	\label{fig:DifferentSigns}
\end{figure}

\subsection{Interaction of compactons with opposite polarities}

Fig.~\ref{fig:DifferentSigns} illustrates the numerical simulation of interaction of two compactons with the same speeds as in Fig.~\ref{fig:SameSigns} but with different polarities. Clearly, the compactons preserve their structures when collide. It follows from the $x-t$ diagram in Fig.~\ref{fig:DifferentSigns}a that the nonlinear shift of coordinate of the slow compacton is opposite to the case of same polarities. During the interaction, both compactons with the opposite polarities experience accelerations. This situation does not occur for the Gardner equation, where the phase shift does not depend on the wave polarity \cite{SlyunyaevPelinovskii1999,Slunyaev2001}. At the same time, a  similar picture of the phase shifts was observed for interacting compactons of the degenerate KdV equation (\ref{kdv}) with $m=3$ and $n=3$ in \cite{RosenauOron2014}.

A large short-living wave occurs during the collision with the amplitude approximately equal to the sum of the compacton amplitudes $|A_1| + |A_2| \approx 2.0$, the maximum wave possesses strong vertical asymmetry, see  Figs.~\ref{fig:DifferentSigns}b,c. The shape of the maximum wave (Fig.~\ref{fig:DifferentSigns}d, $t = 88$) is typical for the extreme events which occur as a result of absorb-emit interaction of solitons of different sings in the modified KdV equation \cite{Slunyaev2001} (see also \cite{Zhangetal2014,PelinovskyShurgalina2015,Shurgalina}). However, 
the sign of the extreme wave is the same as the sign of the slower (larger) compacton. This feature is different from the Gardner equation, where the sign of the amplified wave is specified by the faster soliton \cite{Slunyaev2019}.

The collision of compactons of opposite signs occurs inelastically. A development of a train of positive compactons is readily seen in Fig.~\ref{fig:DifferentSigns}e at $t = 97$, which displays the area to the left from the interacting compactons. The estimated parameters for two of them are plotted in Fig.~\ref{fig:SolitaryWaveParameters} with crosses; they corresponds to compactons. A similar scenario of a collision of two compactons with different signs was observed in \cite{RosenauOron2014} within the framework of the degenerate KdV equation (\ref{kdv}). Negative small-amplitude compactons may also be seen at $t = 100$ in between positive small-amplitude compactons. The given initially compactons approximately restore their amplitudes after the collisions, but the presence of small-amplitude compactons result in irregular oscillations of the wave extremes in time.

\section{Conclusion}

Compactons are typically considered in the systems where the advection and dispersion terms in the corresponding evolution equation are both nonlinear. If the dispersion is linear, then the solitary waves are generally described by hyperbolic functions on the line which decay to zero exponentially at infinity. The generalized KdV equation is not integrable with the exception of the quadratic and cubic powers and their linear combinations (the Gardner equation). 

Here we have studied compactons in the sublinear KdV equation (\ref{kdv-sl}). These compactons are larger in amplitude and wider when the speed is smaller. They may be either positive or negative, and propagate to the  same direction as the dispersive waves of the linear KdV equation. We have shown that the compactons are energetically stable with respect to symmetric compact perturbations with the same support.

The key elements of the compacton dynamics have been studied numerically in this work for the particular case of the sublinear nonlinearity, $\alpha =3/4$. The stability of compactons is confirmed in the direct numerical simulations of propagating and interacting compactons. The long-term solution of the evolution problem for pulse-like disturbances is shown to tend to a sequence of compactons. Due to the specific relation between their amplitudes and speeds, the trains of small-amplitude compactons are ordered differently to the case of the integrable KdV equation. Compactons with the opposite polarity may arise in the time evolution of the pulse-like initial data.

Collisions of compactons are not elastic. Interacting compactons emit waves in the form of trains of small-scale compactons, but almost recover their amplitudes after collisions, so that they may persist during numerous collisions with small or large compactons. The three types of collisions known for solitons have been observed in the present framework: overtaking and exchange interaction of compactons with the same polarity, and absorb-emit interaction of compactons with the opposite polarities. Similar to solitons of the Gardner equation, the maximum wave field decreases when compactons of the same polarity interact, or increases in the case of opposite polarities. Compactons with the amplitudes $A_1$ and $A_2$ result in a wave with the amplitude given by the difference of amplitudes $A_1$ and $A_2$ when they merge, where the amplitudes may have either signs.
Interacting compactons experience nonlinear shifts of coordinates. The dynamics of sublinear compactons demonstates much similarity with other compacton-carrying equations.

Compactons play a twofold role in the wave evolution within the sublinear KdV equation. On the one hand, compactons behave similar to solitary waves: they survive in collisions with other waves and represent the long-term solution of the evolution problem. On the other hand, compactons play the role of dispersive waves of the linear KdV equation, when they quickly spread the residual energy of the initial perturbation which has not been taken by large-amplitude compactons. In the first role, a slowly decaying smooth tail appears first from the left of the perturbation, which splits later on into small-amplitude compactons. In the second role, new small-amplitude compactons are emitted by inelastically interacting compactons.

\vspace{0.25cm}

{\bf Acknowledgements.} The authors thank P. Rosenau and the anonymous reviewers for useful suggestions 
which helped them to improve the manuscript. 
The numerical simulations (ASl) were supported by Laboratory of Dynamical Systems and Applications NRU HSE, of the Ministry of science and higher education of the RF grant ag. No 075-15-2019-1931.
The other study is supported by the Russian Science Foundation under the grant No. 19-12-00253.

\end{document}